\documentclass[12pt]{spieman}  
\usepackage{amsmath,amsfonts,amssymb}
\usepackage{graphicx}
\usepackage{setspace}
\usepackage{tocloft}
\usepackage[utf8]{inputenc}
\usepackage{siunitx}
\usepackage[table]{xcolor}
\usepackage{booktabs}
\usepackage{multirow}

\title{Retinal analysis of a mouse model of Alzheimer's disease with multi-contrast optical coherence tomography}

\author[a,*]{Danielle J. Harper}
\author[a]{Marco Augustin}
\author[a]{Antonia Lichtenegger}
\author[a,b]{Johanna Gesperger}
\author[c]{Tanja Himmel}
\author[a]{Martina Muck}
\author[a]{Conrad W. Merkle}
\author[a]{Pablo Eugui}
\author[d]{Stefan Kummer}
\author[b]{Adelheid Woehrer}
\author[d,$\dagger$]{Martin Glösmann}
\author[a,$\dagger$]{Bernhard Baumann}

\affil[a]{Center for Medical Physics and Biomedical Engineering, Medical University of Vienna, Waehringer Guertel 18-20/4L, 1090 Vienna, Austria}
\affil[b]{Institute of Neurology, General Hospital and Medical University of Vienna, Waehringer Guertel 18-20/4J, 1090 Vienna, Austria}
\affil[c]{Institute of Pathology, University of Veterinary Medicine, Vienna, Veterinaerplatz 1, 1210 Vienna, Austria}
\affil[d]{Core Facility for Research and Technology, University of Veterinary Medicine, Vienna, Veterinaerplatz 1, 1210 Vienna, Austria}
\affil[$\dagger$]{These authors jointly directed this work.}

\cftpagenumbersoff{figure}
\cftpagenumbersoff{table} 
\begin{document} 

\maketitle
\definecolor{TGblue}{rgb}{0.44, 0.5, 0.73}
\definecolor{WTred}{rgb}{1, 0.398, 0.398}
{\noindent \footnotesize\textbf{*}Danielle J. Harper,  \linkable{danielle.harper@meduniwien.ac.at}}

\begin{abstract}
Recent Alzheimer’s disease (AD) patient studies have focused on retinal analysis, as the retina is the only part of the central nervous system which can be imaged non-invasively by optical methods. However as this is a relatively new approach, the occurrence and role of pathological features such as retinal layer thinning, extracellular amyloid beta (A$\beta$) accumulation and vascular changes is still debated. Animal models of AD are therefore often used in attempts to understand the disease. In this work, both eyes of 24 APP/PS1 transgenic mice (age: 45-104 weeks) and 15 age-matched wildtype littermates were imaged by a custom-built multi-contrast optical coherence tomography (OCT) system. The system provided a combination of standard reflectivity data, polarization-sensitive data and OCT angiograms. This tri-fold contrast provided qualitative and quantitative information on retinal layer thickness and structure, presence of hyper-reflective foci, phase retardation abnormalities and retinal vasculature. While abnormal structural properties and phase retardation signals were observed in the retinas, the observations were very similar in transgenic and control mice. At the end of the experiment, retinas and brains were harvested from a subset of the mice (14 transgenic, 7 age-matched control) in order to compare the in vivo results to histological analysis, and to quantify the cortical A$\beta$ plaque load.
\end{abstract}

\begin{spacing}{1}   

\section{Introduction}
\label{sect:intro}  

It is hypothesized that cerebral changes precede Alzheimer's disease (AD) symptom presentation by over 20 years \cite{Jack2009}. Since the beginning of the millennium, the number of deaths from AD in the United States has increased by 145\%. For comparison, the number of deaths caused by heart disease (the number one cause of death in the United States) has decreased by 9\% in the same time period \cite{alzFacts}. AD is a chronic and irreversible neurodegenerative disorder with no current cure. The time delay between the start of the disease and the presentation of symptoms means that the disease is already at an advanced stage before it can be detected, and even when a patient presents with AD symptoms, a definitive AD diagnosis still remains challenging.

Post mortem diagnosis of AD is realized by the positive histological identification of both extracellular amyloid beta (A$\beta$) plaques and intracellular neurofibrillary tau tangles, which are both found in the brains of AD patients \cite{funato1998quantitation,Hurtado-Puerto2018}. A key step along the road to an early AD diagnosis would be an in vivo identification of A$\beta$ plaques. However as the plaques are small (ranging from 10-200 \si{\micro\meter} \cite{dudeffant2017contrast}) and located within the brain, this presents some logistical difficulties.

One recent idea to circumvent these difficulties is the use of the eye ``as a window to the brain''. The retina and the brain are derived from the same embryological origin; they both extend from the neural tube. The retina is therefore the only part of the central nervous system which can be imaged non-invasively by optical methods. However the question still remains whether the retina can hold the key to early AD diagnosis as there are few studies which directly correlate findings in the retina to those in the brain. Those studies which have been done before have found a correlation between the amplitude of retinal vascular pulsatility and neocortical A$\beta$ scores (measured using florbetaben positron emission tomography) \cite{golzan2017retinal}, and also between fluorescent components (measured using fluorescence lifetime imaging ophthalmoscopy) and both p‐tau181‐protein concentration in the cerebral spine fluid and the mini-mental state examination score \cite{jentsch2015retinal}. 

Recent studies have also focused on the identification of extracellular A$\beta$ accumulations in the retina of AD patients, however there are conflicting reports on this topic \cite{ong2018controversies}. Some reports have identified extracellular A$\beta$ in the retina \cite{koronyo2011identification,tsai2014ocular,la2016melanopsin,koronyo2017retinal}, however other groups have demonstrated that there was no A$\beta$ to be found \cite{blanks1989retinal,schon2012long,ho2014beta,den2018amyloid}. Another topic of current debate is whether AD could also be considered a vascular disorder\cite{de2004alzheimer}, and therefore the retinal vasculature of AD patients has also recently been studied. Patients with AD may exhibit more tortuous retinal vessels \cite{Cheung2014}. Narrowing of retinal blood vessels and reduced venous blood flow rates have also both been found in AD patients \cite{berisha2007retinal,Feke2015}, and an overall more sparse retinal microvascular network has been observed \cite{Williams2015,zabel2019comparison}. A recent OCT angiography (OCTA) study has shown a reduced vessel density specifically in the superficial capillary plexus \cite{yoon2019retinal}. It has also been suggested that blood flow changes may precede neurodegeneration \cite{Feke2015}.

While the aforementioned studies have all demonstrated a reduction in blood circulation in the retina in late stage AD, recent studies have indicated that the retinal vessel density \cite{vandeKreekeARVO} and vessel diameter \cite{LaughlinARVO} both seem to be increased in individuals suffering from preclinical AD. Such results are consistent with the theory that an inflammatory response occurs in the retina in the early stages of AD - a theory which has also been proposed for neurovasculature \cite{akiyama1994inflammatory,heneka2015neuroinflammation}.

Retinal layer thinning, particularly in the retinal nerve fiber layer (RNFL), is also present in the retina of AD patients \cite{den2017retinal,chan2019spectral}. However looking forward to a marker for diagnosis, RNFL thinning is not specific to AD and it is not only associated with other diseases such as glaucoma \cite{leung2012retinal} and Parkinson's disease \cite{inzelberg2004retinal}, but also more generally with increasing age \cite{alasil2013analysis}.

With many contradictory observations, it is clear that there is still a great deal of research to be performed in order to fully understand the effects of AD on the retina. Some attempts to do this have focused on the use of mouse models of the disease. When performing studies on animal models, it is important to know where the similarities and differences to the human disease lie. While there are many mouse models of AD \cite{hall2012mouse}, this work focuses on a doubly transgenic model which expresses a chimeric mouse/human amyloid precursor protein (APP) and a mutant human presenilin 1 (PS1) \cite{Jankowsky2001,Jankowsky2003,Reiserer2007}. The following paragraphs describe the retinal changes observed in this transgenic APP/PS1 mouse model so far.

Much like in the case of the human, while well documented in the brain, \cite{yan2009characterizing,ordonez2016abetapp} the appearance of extracellular deposits of A$\beta$ in the retina is disputed. Several studies have reported no extracellular deposits of A$\beta$ in the retina, despite plaques being present in the brain and an increased expression of APP in the retina, similar to what has been found in humans \cite{ho2014beta}. This has been reported in mice of several ages: 9 months old \cite{Dutescu2009}, 7-12 months old \cite{Shimazawa2008} and 13 months old \cite{Joly2017}. It has been suggested that the nonamyloidogenic pathway may endogenously limit A$\beta$ formation in the retina \cite{Joly2017}. A further extensive histological analysis of the retina concluded that no identifiable retinal pathology exists in these mice \cite{chidlow2017investigations}. Conversely, extracellular deposits of A$\beta$ were found at the age of 27 months old in the choriocapillaris and the nerve fiber layer, but not in the other layers \cite{Ning2008}. In another study, plaques were found in the inner plexiform layer and in the outer plexiform layer. These plaques ranged in size from 5-20 \si{\micro\meter} in mice of 12-13 months old, and larger with increasing age \cite{Perez2009}. This study also reported no observed changes in retinal layer thickness. Deposits were also found distributed throughout the retina of transgenic mice at the age of 9 months and 17 months, but were identifiable as young as 2.5 months, even before the plaques appeared in the brain \cite{koronyo2011identification}. 

There is therefore a need for more studies linking the retina and the brain in both AD patients and in animal models of AD \cite{Chiquita2019}, and more work needs to be done to assess and quantify the presence of A$\beta$ in the retina \cite{shah2017beta}. Optical coherence tomography (OCT) \cite{huang1991optical} may be a useful tool to employ for this purpose. As a non-contact, non-invasive imaging modality, OCT has become part of clinical routine for in vivo retinal diagnostics. Functional extensions of OCT have made it possible to not only visualize contrasts based on backscattered intensity (reflectivity), but also motion (OCTA) \cite{de2015review,zhu2017can,chen2017optical} and polarization properties (PS-OCT) such as birefringence \cite{hee1992polarization,pircher2011polarization,de2017polarization,baumann2017polarization}. The birefringence of A$\beta$ plaques has been studied in detail using polarimetry \cite{jin2003imaging,campbell2015polarization,hamel2016polarization,campbell2018amyloid}, and also with PS-OCT \cite{Baumann2017,gesperger2019comparison}. The plaques appear as hyper-scattering structures in standard reflectivity OCT images \cite{bolmont2012label,Marchand2017,Lichtenegger2018}, but the addition of polarization-sensitive detection provides an additional tissue-specific contrast. Furthermore, the combination of PS-OCT with OCTA allows a simultaneous analysis of this tissue-specific contrast and changes in the retinal vasculature.  

In this work, the appearance of the retina of an APP/PS1 mouse model of AD was evaluated by multi-contrast spectral domain OCT. By observing retinal changes over a range of ages (45-104 weeks) in \hbox{intensity-,} polarization- and motion-based contrast modes, the observations in the retina were documented, mapped directly to histology, and compared to the A$\beta$ plaque load in the brain.   

\section{Materials and Methods}

\subsection{Optical coherence tomography}
\label{sect:OCT}

A modified version of a PS-OCT system described elsewhere was used in this study \cite{fialova2016polarization}. In brief, the system operated at a central wavelength of 840 \si{\nano\meter} with a full width at half maximum bandwidth of approximately 100 \si{\nano\meter}, resulting in an axial resolution of around 3.8 \si{\micro\meter} in retinal tissue. Light incident upon the mouse eye was of a known polarization state, and the polarization-sensitive detection allowed for the differentiation between polarization-preserving tissue and polarization-altering tissue. 

\begin{figure}
\begin{center}
\begin{tabular}{c}
\includegraphics[height=8.5 cm]{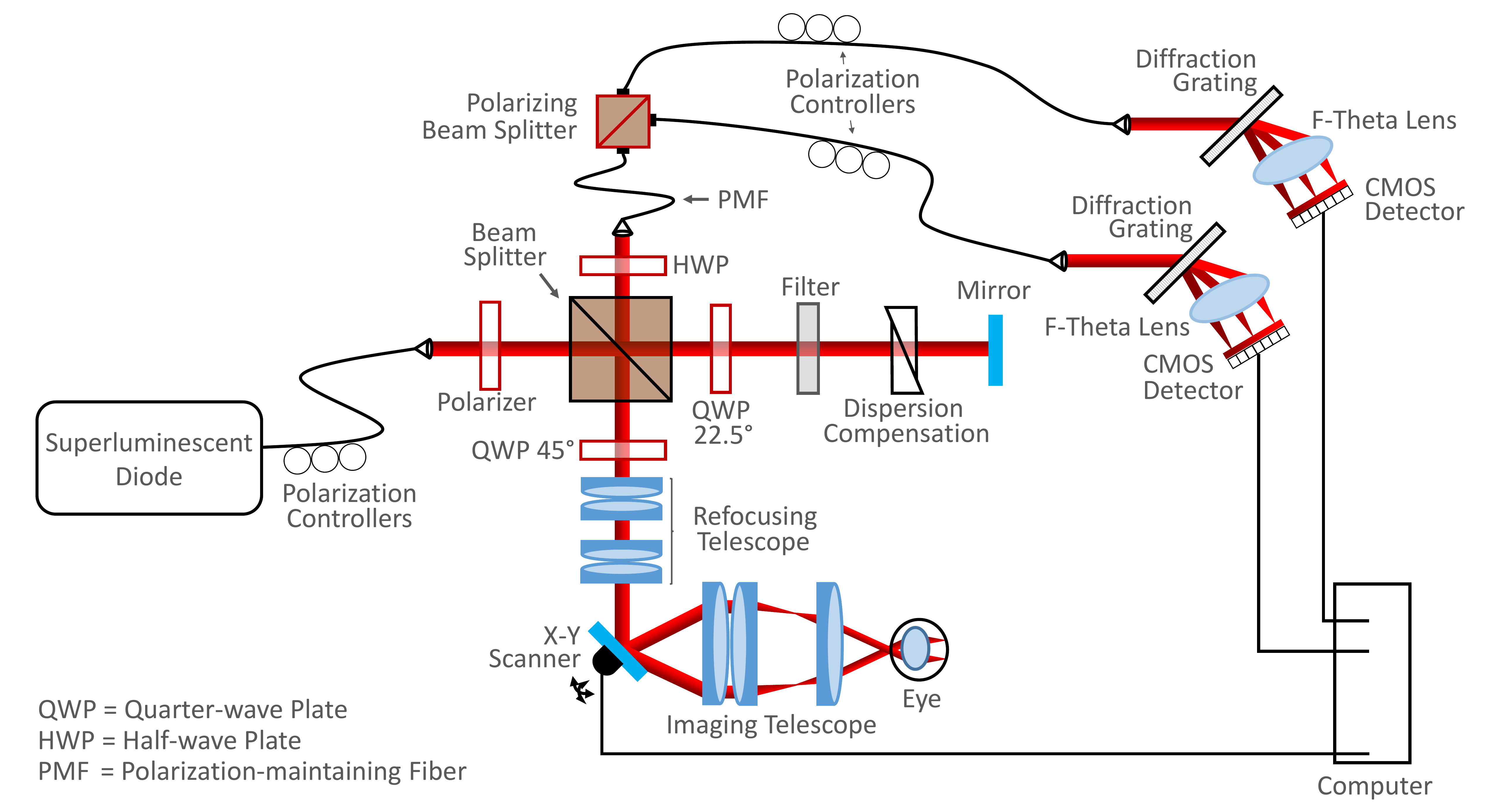}
\end{tabular}
\end{center}
\caption 
{ \label{fig1}
A modified version of the PS-OCT system first described by Fialová et al. \cite{fialova2016polarization}. A refocusing telescope was added to the system to allow focus correction of each individual mouse eye.} 
\end{figure}

An additional refocusing telescope was added to the system to correct for aberrations induced by poor focus of the mouse eye itself \cite{harper2018white}. A diagram of the modified version of the system can be found in Fig. \ref{fig1}. Two additional achromatic doublet pairs (2 $\times$ AC254-080-B, Thorlabs and 2 $\times$ AC254-050-B, Thorlabs) were mounted on a translational stage, allowing the focus to be manually optimized for each individual mouse eye while reducing the beam diameter incident on the pupil from 0.8 \si{\milli\meter} to 0.5 \si{\milli\meter}. Each mouse eye was aligned with respect to the 2.85 \si{\milli\watt} measurement beam to ensure the optic nerve head (ONH) was at the center of the 1 \si{\milli\meter} $\times$ 1 \si{\milli\meter} field of view. With an A-scan rate of 83 \si{\kilo\hertz}, five repeated B-scans (consisting of 512 A-scans each) were acquired at 400 unique locations. Such a scan pattern allowed for an increased signal-to-noise ratio (SNR) in the reflectivity and PS-OCT images, and also the ability to produce OCTA images.

\subsection{Mice}
\label{sect:mice}

A breeding pair of APP/PS1 mice (APPswe, PSEN1dE9 MMRRC stock number 34829-JAX) was purchased from The Jackson Laboratory (Bar Harbor, ME, USA) \cite{Jankowsky2001,Jankowsky2003,Reiserer2007}, and a breeding colony was established at the Division of Biomedical Research at the Medical University of Vienna. Hemizygous mice were bred with wildtype siblings and subsequently kept under controlled lighting conditions (12 hours light, 12 hours dark) with food and water ad libitum. Both eyes of 24 mutant mice (17 female, 7 male) and 15 wild-type littermates (9 female, 6 male) were imaged using the multi-contrast OCT system. At the time of imaging, the mice ranged in age from 45 weeks to 104 weeks. During the experiment, the animals were anesthetized using an inhalational isoflurane/oxygen mixture (4\% isoflurane for 4 minutes in an induction chamber to induce anesthesia, 2\% delivered via a nose cone thereafter). To facilitate the OCT imaging, pupils were dilated using topically applied tropicamide and phenylephrine. The cornea was kept moisturized using artificial tear eye drops, and heating pads were placed underneath the mice to prevent a reduction in body temperature. All experiments were performed in accordance with the ARVO Statement for the Use of Animals in Ophthalmic and Vision Research and Directive 2010/63/EU. Ethics protocols were approved by the ethics committee of the Medical University of Vienna and the Austrian Federal Ministry of Education, Science and Research (BMBWF/66.009/0272-V/3b/2019). 

\begin{figure}
\begin{center}
\begin{tabular}{c}
\includegraphics[height=6.1 cm]{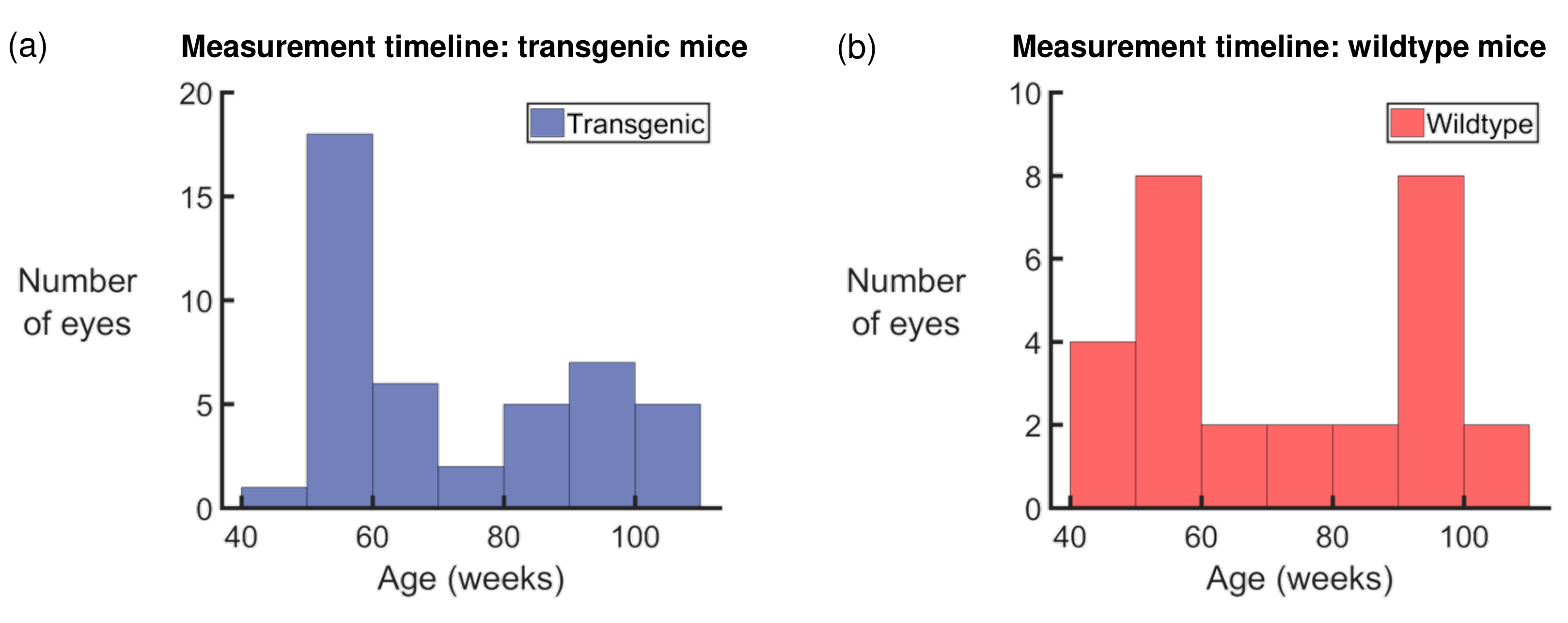}
\end{tabular}
\end{center}
\caption 
{ \label{fig2}
Histogram representations of the number of eyes used for analysis. A total of 44 eyes from 24 APP/PS1 transgenic mice (a) and 28 eyes from 15 wildtype littermates (b) were imaged within an age range of 45 weeks to 104 weeks.} 
\end{figure} 

\subsection{OCT image analysis}
\label{sect:analysis}
\begin{figure}
\begin{center}
\begin{tabular}{c}
\includegraphics[height=13.0cm]{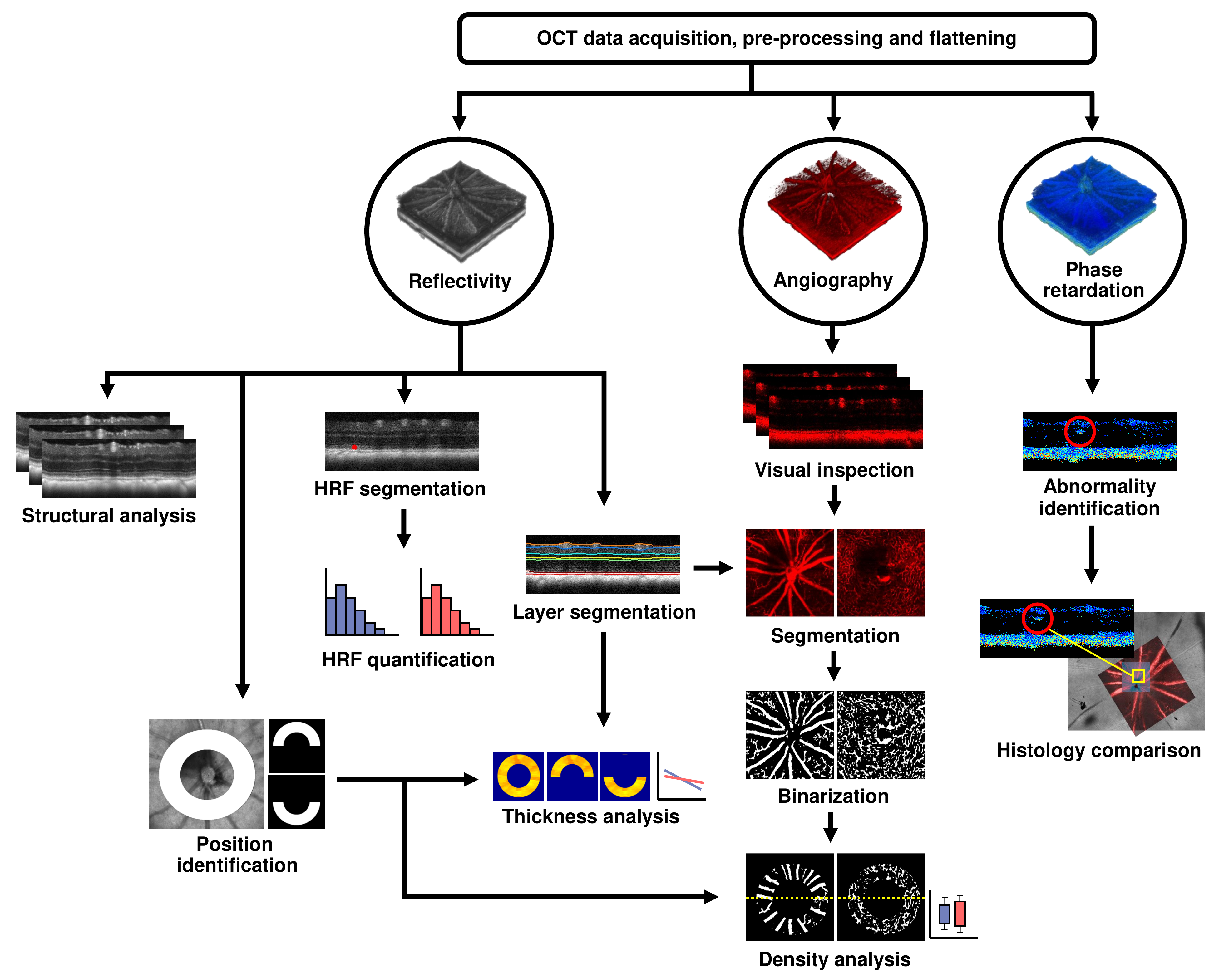}
\end{tabular}
\end{center}
\caption 
{ \label{fig3}
A flow chart of the multi-contrast optical coherence tomography (OCT) post-processing pipeline which consists primarily of reflectivity, phase retardation and angiography data. \textbf{HRF} Hyperreflective foci.} 
\end{figure}

The analysis performed in this work was based on a previously-published multi-contrast image processing pipeline including standard intensity-based reflectivity contrast, polarization-based contrast and motion-based angiographic contrast \cite{augustin2016multi}. Prior to analysis, the images were corrected for axial motion and the retina was flattened with respect to the retinal pigment epithelium (RPE)/choroidal complex as detected by the cross-polarized channel \cite{augustin2018segmentation}, a technique made possible by the polarization sensitive detection. Any dataset for which the retinal flattening failed (due to poor signal-to-noise ratio) was excluded, resulting in a total of 72 datasets for evaluation (44 eyes from 24 transgenic mice and 28 eyes from 15 wildtype control mice). A graphical visualization of the age of the mice at each measurement can be found in Fig. \ref{fig2}. A flow chart of the overall post-processing pipeline can be seen in Fig. \ref{fig3}, and a description of the analysis can be found in the following sections.

\subsubsection{Retinal thickness analysis}

To allow for a comprehensive retinal thickness analysis, layer segmentation of the retina was first performed using a previously-described algorithm \cite{augustin2018segmentation}. The distance between the inner limiting membrane (ILM) and the posterior surface of the RPE was defined as the total retinal thickness. The posterior surface of the outer plexiform layer (OPL) served as the boundary between the inner and outer retina, as defined in this manuscript. An annulus around the optic nerve head was created with an inner diameter of 500 \si{\micro\meter} and an outer diameter of \hbox{900 \si{\micro\meter},} and the mean total, inner and outer retinal thicknesses were evaluated within this annular 3D volume. The annulus was then cut in half transversely and the mean thicknesses were again calculated in the two resulting regions, corresponding to the superior and inferior retina. Due to mouse eye alignment, it is estimated that this boundary line is accurate to within $\pm$ 30 \si{\degree}. Coordinates of the segmentation lines and the ONH annuli were stored for later use in OCTA analysis.

To test for statistical significance, the retinal thickness measurements were plotted as a function of age for both the transgenic and the wildtype mice. A linear regression pre-test was performed to determine if the thickness measurements were dependent on age. If the results were deemed significant ($p < 0.05$), an analysis of covariance (ANCOVA) was performed to test for a difference in the trend between the two groups. In the case where the pre-test was deemed statistically insignificant ($p \geq 0.05$), regular analysis of variance (ANOVA) was performed to test for a difference in the group means. The gradient of the slope of the regression lines were also documented for both transgenic and wildtype mice, corresponding to a measurement of reduction of retinal thickness in units of \si{\micro\meter} per week.  

\subsubsection{Hyperreflective foci analysis}

All 72 reflectivity datasets (44 eyes from 24 transgenic mice and 28 eyes from 15 wildtype control mice) were manually screened for hyperreflective foci in the posterior layers of the retina, spanning the region from the posterior OPL border to the posterior RPE (i.e. the whole outer retina). The hyperreflective foci were then manually segmented using ITK Snap \cite{yushkevich2006user}. Using the data from this segmentation, the number and location of hyperreflective foci was evaluated for each eye. 

\subsubsection{Polarization properties}

The phase retardation images were calculated for every B-scan by 

\begin{equation} \label{retardation}
\delta = \arctan\left(\dfrac{A_V}{A_H}\right) 
\end{equation} 

\noindent where $A_V$ and $A_H$ correspond to the signal amplitudes of the co- and cross-polarized channels, respectively \cite{hee1992polarization,gotzinger2005high}. In the healthy mouse retina, high retardation values are only expected in the melanin-containing regions, namely the RPE, the choroid, and the remnants of the hyaloid artery near the ONH. Depolarization occurs due to the fact that the melanin granules scramble the polarization state of the incoming light beam, resulting in random phase retardation values. Polarization-preserving tissues, i.e., the rest of the retina, do not retard the phase of the incident beam, and therefore the phase retardation values are low. All retardation B-scans were manually inspected for abnormally high retardation signals from outwith these retinal layers. The number of these abnormally high retardation signals was evaluated, and the sources were investigated with retinal histology. 

\subsubsection{OCT angiography}

The B-scan repetition allowed for the computation of OCTA images, revealing locations of motion contrast. After bulk motion compensation and removal of frames with uncorrectable motion, OCTA images were computed by calculating the averaged magnitude of the complex differences between consecutive repeated B-scans. The time delay between the acquisition of repeated B-scans ($\approx$ 7.7 \si{\milli\second} from one B-scan to the next) which provides the angiographic contrast in the first place also makes the method very susceptible to motion. All datasets were therefore manually visually screened, and data which included severe motion artefacts or regions of poor angiography signal were excluded. The angiography analysis was therefore performed on 39 transgenic eyes and 16 wildtype eyes. 

An automated OCTA processing pipeline was created which consisted of several steps. The superficial vascular plexus (SVP) and the deep capillary plexus (DCP) were segmented from the retina using the layer segmentation coordinates obtained from the reflectivity data (SVP corresponds to the RNFL; DCP corresponds to the OPL). A maximum intensity en-face projection was then calculated over the SVP and the DCP independently, and the histograms of each image were equalized using contrast limited adaptive histogram equalization \cite{zuiderveld1994contrast} before image binarization. The binary images were then morphologically opened and closed (disk-shaped structuring element with a radius of 1), and skeletonized to remove speckle noise and enhance the vessel connections. A square averaging filter with a 5-pixel side length was then applied to the images to create the final binary vessel representations of ``vessel'' vs. ``non-vessel''. The annuli around the ONH which had been calculated using the reflectivity data were then applied to these vessel maps. The vessel density was then calculated as the percentage of pixels which were marked as ``vessel'' in the whole annulus, as well as in the superior and inferior retinal regions.

After this image processing was performed, a modified version of the Weber contrast \cite{weber} was calculated to test the relationship of the OCTA vessel intensity to the background, using the binary image as a mask. The mean intensity value of all pixels determined both ``vessels'' ($\overline{I_{v}}$), and ``non-vessels'' ($\overline{I_{b}}$), was calculated, and the modified Weber contrast, $C_W$ was calculated as

\begin{equation}
C_W = \frac{\overline{I_{v}}-\overline{I_{b}}}{\overline{I_{b}}}
\end{equation}

\noindent and the results were plotted as box-and-whisker diagrams for both transgenic and wildtype mice.

\subsection{Histology and immunostaining}
\label{sect:methodsANIMALS}

After OCT imaging, a sub-group of the mice (14 transgenic, age: 54-104 weeks, 7 wildtype control, age: 54-103 weeks) were euthanized by cervical dislocation. Immediately after sacrifice, the brains were extracted and the eyes were enucleated for histological analysis. 

\subsubsection{Brain}
The mouse brains were sagittally cut into two hemispheres, and one hemisphere was prepared for histopathological workup. The samples were fixed in 4\% formalin and processed through graded alcohols and xylene into paraffin. Sagittal brain sections with a thickness of 2.5 \si{\micro\meter} were cut on a microtome, deparaffinized, rehydrated, and stained immunohistochemically using an anti-A$\beta$ antibody (clone 6F/3D, diluted 1:100, Dako). The sections were evaluated using a slide scanner (Hamamatsu NanoZoomer 2.0 HT) and saved for digital pathology. The images were analyzed using Fiji \cite{schindelin2012fiji}.  First, the cortex was manually selected and the ``ColSeg" tool \cite{loo2012effects} was utilized to segment the plaques by their brown color. The ``analyze particle" tool was then used to count the plaque number and calculate the plaque load in plaques per $\si{\milli\meter\squared}$. The plaque load was then plotted as a function of mouse age, and a linear regression analysis was performed on the data.

\subsubsection{Retina}
\label{retina}
To obtain vertical histological slices of the retina, six left eyes from wildtype (n=3) and transgenic (n=3) mice diagnosed with and without abnormal OPL banding (see Section \ref{opl}) were immersed unopened in Davidson's fixative for 24 hours at \ang{4}C and processed through graded alcohols and xylene into paraffin. Three-micron-thick sections were then cut, mounted onto slides, deparaffinized, rehydrated, and stained with hematoxylin and eosin (H\&E). 

For wholemount preparation and A$\beta$ immunostaining, eyes were immersed in 4\% paraformaldehyde (PFA) in 0.1 M phosphate buffered saline (PBS), pH 7.4. Some eyes were fixed unopened. In the others, cornea and lens were removed and the eyecups fixed in PFA for at least 24 h at room temperature. After rinses in PBS, the retina was dissected free from RPE, choroid and sclera, cryoprotected in ascending sucrose concentrations (10\%, 20\%, 30\%), and snap-frozen and thawed three times to increase antibody penetration. From each mouse, the left retina was treated with 70\% formic acid for 10 minutes and then rinsed repeatedly in PBS, while the right retina was left untreated. Retinal wholemounts were processed free-floating in 24-well plates and all incubations and rinses were done with gentle rotation on a rocker table at \ang{4}C. Blocking of non-specific binding was performed in 3\% normal donkey serum in 0.1 M PBS, 0.25\% Triton X-100 and 0.05\% sodium azide (medium), followed by incubation with mouse anti-human A$\beta$ (Abcam, ab11132, clone DE2B4, 1:400 in medium) for 72 h. After washes in PBS, retinas were incubated in donkey anti-mouse Fab fragments conjugated with Alexa Fluor 488 (Jackson ImmunoResearch Laboratories, 1:500 in medium) for 24 h, rinsed, and coverslipped (retinas ganglion cell side up) in Aqua/Polymount (Polysciences). To serve as positive and negative controls, respectively, brains from transgenic APP/PS1 mice and their wildtype littermates were harvested after enucleation and fixed in 4\% PFA for 24 h at \ang{4}C. After washes in PBS, brains were cryoprotected in ascending sucrose concentrations (10\%, 20\%, 30\%), snap-frozen in liquid nitrogen-prechilled isopentane and cut into 100-\si{\micro\meter}-thick sections using a cryotome. The sections were collected in PBS/0.05\% sodium azide and processed under the same conditions applied to retinal wholemounts. H\&E-stained sections were examined with brightfield illumination on a Zeiss Axio Imager Z2. Immunofluorescence analysis was performed with a Zeiss LSM880 laser scanning microscope (LSM). A total of 17 retinas from 11 transgenic mice (age: 54-103 weeks) and 11 retinas from 6 wildtype control mice (age: 54-103 weeks) were suitable for detailed histological examination. A 1:1 correlation of the retinal wholemounts to the OCT image data was then performed by mapping the vessel pattern of the SVP as visualized by the LSM to the corresponding OCTA datasets.

\section{Results}
\label{sect:results}

\subsection{Hyperreflective foci}

\begin{figure}
\begin{center}
\begin{tabular}{c}
\includegraphics[height=11.2cm]{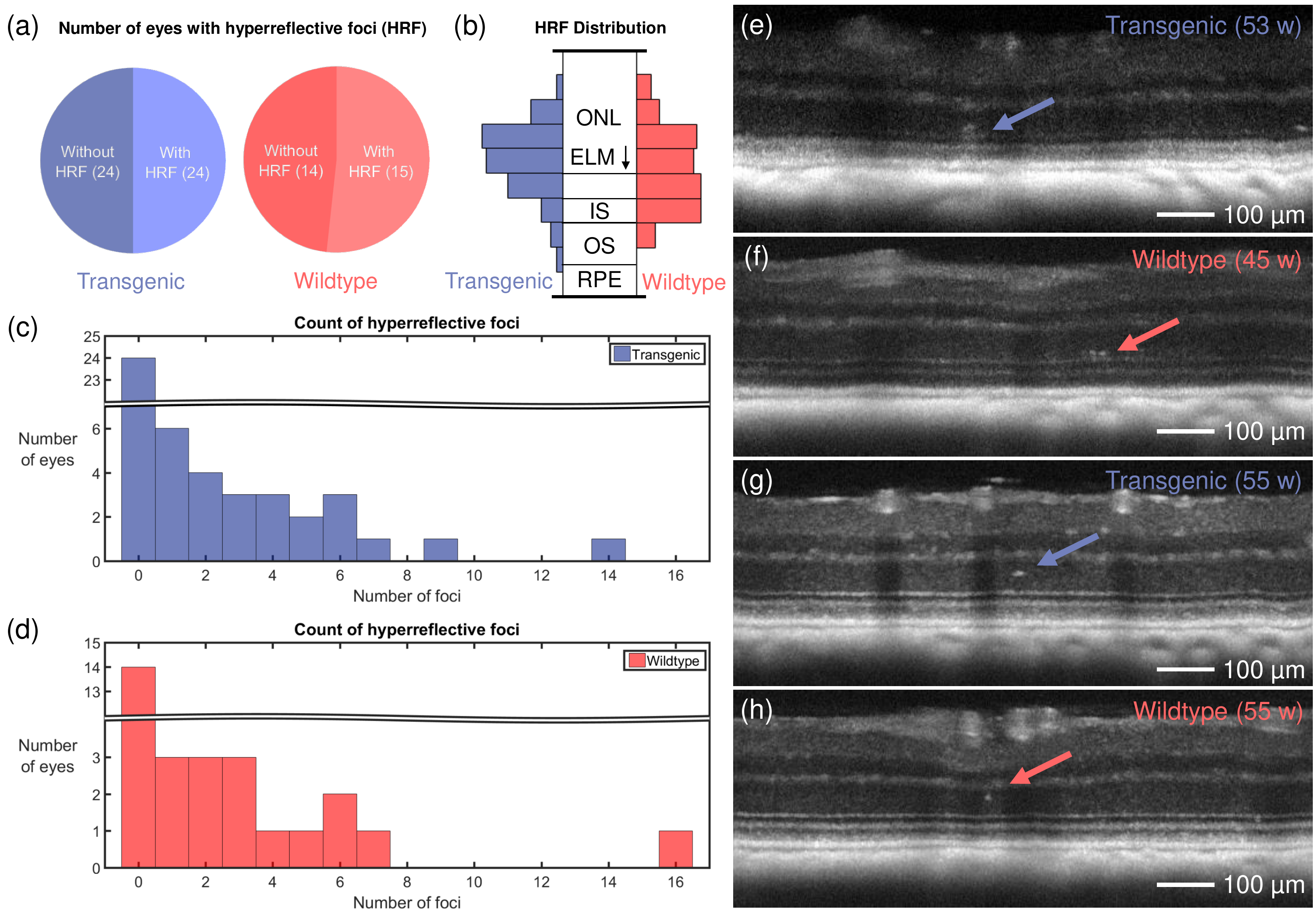}
\end{tabular}
\end{center}
\caption 
{ \label{fig4}
Results of hyperreflective foci (HRF) analysis. a) Pie charts indicating the number of eyes with and without HRF for both transgenic and wildtype mice. b) HRF probability distribution displayed with respect to outer retinal layer position for transgenic and wildtype mice. The distributions are very similar, with most HRF appearing near the external limiting membrane (ELM). \textbf{ONL} outer nuclear layer. \textbf{IS} inner segments. \textbf{OS} outer segments. \textbf{RPE} retinal pigment epithelium. c) Histogram of HRF occurrence in transgenic mice. d) Histogram of HRF occurrence in wildtype mice. e-h) Some examples of the appearance of HRF in OCT reflectivity images. Each image is a maximum intensity projection over four consecutive B-scans, where each B-scan is already averaged 5 times and plotted on a log scale. e-f) HRF located above the ELM in both the transgenic mouse retina (e) and in the wildtype retina (f). g-h) HRF located in the middle of the ONL in both the transgenic mouse retina (g) and in the wildtype retina (h). Ages of mice in weeks (w) are indicated on sub-figures (e-h).} 
\end{figure}

For each retina, the whole $1 \times 1$ \si{\milli\meter\squared} area surrounding the optic nerve head was evaluated. Of the 24 mutant mice, 16 showed HRF in at least one eye. In the wildtype littermate control group, HRF were identified in 12 out of the 15 mice. Figure \ref{fig4}a displays pie charts which document this in terms of eyes; there were an equal number of eyes with and without HRF in the transgenic mice, and a difference of only one in the wildtype mice. Since it is difficult to identify small HRF in the plexiform layers due to the appearance of hyperreflective blood vessels, a normalized probability distribution of all identified HRF in the outer retina alone was plotted (Fig. \ref{fig4}b). A similar HRF distribution in transgenic and wildtype retinas was observed. The number of HRF per eye was also counted for the transgenic (Fig. \ref{fig4}c) and the wildtype (Fig. \ref{fig4}d) mice. With the exception of one outlier in each group, all outer retinas contained less than 10 HRF within the investigated field of view. Qualitatively, the types of HRF also looked very similar between transgenic and wildtype mice, examples of which can be found in Fig. \ref{fig4}e-h. Figure \ref{fig4}e and Fig. \ref{fig4}f show examples of larger HRF located anterior to the ELM in the transgenic and wildtype animals, respectively, while Fig. \ref{fig4}g and Fig. \ref{fig4}h show smaller HRF in the middle of the ONL. Neither the number of HRF nor the size correlated with the age of the mice, for either group.

\subsection{Retinal thickness}

\begin{figure}
\begin{center}
\begin{tabular}{c}
\includegraphics[height=15.3cm]{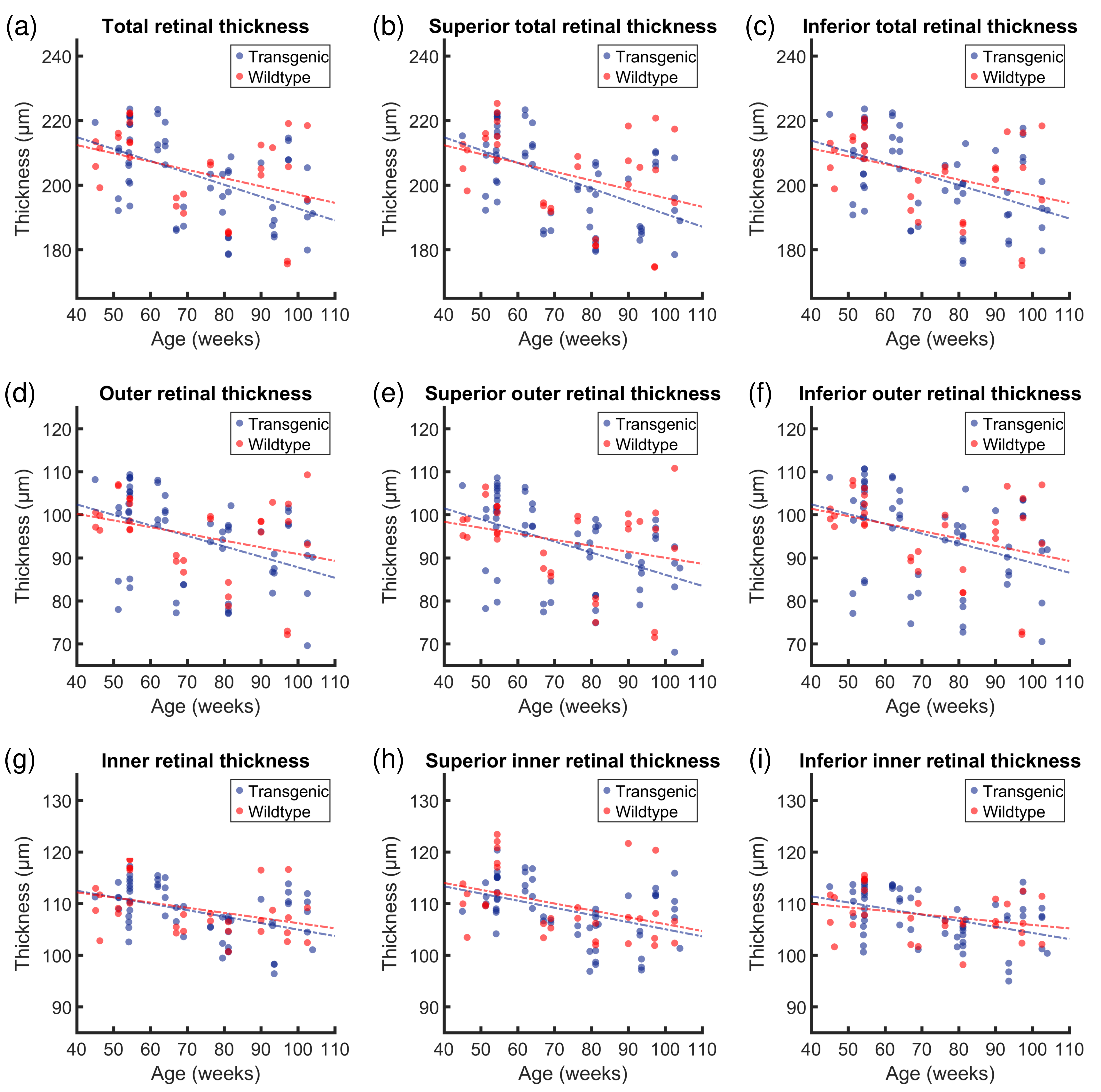}
\end{tabular}
\end{center}
\caption 
{ \label{fig5}
Analysis of retinal thickness as a function of age for both transgenic and wildtype mice. a-c) Total retinal thickness measured around the whole annulus (a), then subdivided into a superior half (b) and an inferior half (c). d-f) Outer retinal thickness measured around the whole annulus (d), in the superior half (e) and in the inferior half (f). g-i) Inner retinal thickness measured around the whole annulus (g), in the superior half (h) and in the inferior half (i). The corresponding statistical evaluation can be found in Table \ref{thickness_stats}, and the gradients of the slopes can be found in Table \ref{slopes}.} 
\end{figure}

The total, outer and inner retinal thickness was calculated in an annulus around the ONH, and also for the superior and inferior ($180 \pm 30$) degree sectors, resulting in nine thickness comparisons between transgenic and wildtype mice. The results of this analysis can be found in Fig. \ref{fig5}. When plotted as a function of age, all nine datasets displayed a general trend of decreasing retinal thickness with age for both the transgenic mice and the wildtype controls. Statistical pre-tests revealed that the dependence upon age was significant for all data except three wildtype datasets: the total outer retinal thickness ($p=0.075$), the superior outer retinal thickness ($p=0.124$) and the inferior inner retinal thickness ($p=0.101$). For these three datasets, the comparison between transgenic and wildtype data were analyzed with ANOVA, while all other analysis was performed with ANCOVA. No statistical significance was found between the retinal thickness changes in the transgenic and control groups, for any of the retinal regions. The results of the statistical analysis can be found in Table \ref{thickness_stats}. Using the gradient of the slope of the linear regression analysis, a measurement of the decrease of retinal thickness was documented in units of \si{\micro\meter} per week, as shown in Table \ref{slopes}. All trends were negative, therefore the gradient of each slope is the negative of the value in the table.

\begin{table}[ht]
\caption{Statistical evaluation for the retinal thickness analysis. All values displayed are p-values, and significance is defined as $p< 0.05$. \textcolor{TGblue}{$\blacksquare$} indicates pre-test p-values for retinal thickness as a function of age for the transgenic group. \textcolor{WTred}{$\blacksquare$} indicates pre-test p-values for retinal thickness as a function of age for the wildtype group. Pre-test p-values were calculated using linear regression analysis. \textcolor{TGblue}{$\diagdown$} \hspace{-8pt}\textcolor{WTred}{$\diagdown$} indicates p-values from ANCOVA comparing the trends of the transgenic retinal thickness vs. age to the wildtype retinal thickness vs. age. \textcolor{TGblue}{$\diamond$}\textcolor{WTred}{$\diamond$} indicates values where the initial pre-tests failed, and therefore one-way ANOVA was performed to test for significance between the means of the two groups. There were no statistically significant changes in retinal thickness between the transgenic and the wildtype groups.} 
\label{thickness_stats}
\begin{center}      
{\renewcommand{\arraystretch}{1.5}
\begin{tabular}{c|c|c|c|c|c|c|c|}
\multicolumn{2}{c}{} & \multicolumn{6}{c}{Position} \\ \cline{3-8} 
\multicolumn{2}{c|}{\multirow{-2}{*}{}} & \multicolumn{2}{c|}{Total} & \multicolumn{2}{c|}{Superior} & \multicolumn{2}{c|}{Inferior} \\ \cline{2-8} \cline{3-8} 
&  & \textcolor{TGblue}{$\blacksquare$} $6.146\times 10^{-5}$ &  &\textcolor{TGblue}{$\blacksquare$} $1.598\times 10^{-5}$ &  &\textcolor{TGblue}{$\blacksquare$} $2.858\times10^{-4}$ &  \\ \cline{3-3} \cline{5-5} \cline{7-7}
& \multirow{-2}{*}{\rotatebox{90}{Total\hspace{8pt}}} & \textcolor{WTred}{$\blacksquare$} $0.033$ & \multirow{-2}{*}{\textcolor{TGblue}{$\diagdown$} \hspace{-10pt}\textcolor{WTred}{$\diagdown$} $0.421$} & \textcolor{WTred}{$\blacksquare$} $0.037$ & \multirow{-2}{*}{\textcolor{TGblue}{$\diagdown$} \hspace{-10pt}\textcolor{WTred}{$\diagdown$} $0.392$} & \textcolor{WTred}{$\blacksquare$} $0.034$ & \multirow{-2}{*}{\textcolor{TGblue}{$\diagdown$} \hspace{-10pt}\textcolor{WTred}{$\diagdown$} $0.466$} \\ \cline{2-8} 
&  & \textcolor{TGblue}{$\blacksquare$} $9.010\times10^{-4}$ &  &\textcolor{TGblue}{$\blacksquare$} $2.159\times10^{-4}$ &   &\textcolor{TGblue}{$\blacksquare$} $0.004$ & \\ \cline{3-3} \cline{5-5} \cline{7-7}
& \multirow{-2}{*}{\rotatebox{90}{Outer\hspace{8pt}}} &\textcolor{WTred}{$\blacksquare$} $0.075$ & \multirow{-2}{*}{\textcolor{TGblue}{$\diamond$}\textcolor{WTred}{$\diamond$} $0.727$} & \textcolor{WTred}{$\blacksquare$} $0.124$ & \multirow{-2}{*}{\textcolor{TGblue}{$\diamond$}\textcolor{WTred}{$\diamond$} $0.716$} & \textcolor{WTred}{$\blacksquare$} $0.045$ & \multirow{-2}{*}{\textcolor{TGblue}{$\diagdown$} \hspace{-10pt}\textcolor{WTred}{$\diagdown$} $0.647$} \\ \cline{2-8} 
& &\textcolor{TGblue}{$\blacksquare$} $3.998\times10^{-4}$ & &\textcolor{TGblue}{$\blacksquare$} $5.437\times10^{-4}$ &  &\textcolor{TGblue}{$\blacksquare$} $7.544\times10^{-4}$ &  \\ \cline{3-3} \cline{5-5} \cline{7-7}
\multirow{-6}{*}{\rotatebox{90}{Retinal Thickness}} & \multirow{-2}{*}{\rotatebox{90}{Inner\hspace{8pt}}} & \textcolor{WTred}{$\blacksquare$} $0.038$ & \multirow{-2}{*}{\textcolor{TGblue}{$\diagdown$} \hspace{-10pt}\textcolor{WTred}{$\diagdown$} $0.640$} &\textcolor{WTred}{$\blacksquare$} $0.030$ & \multirow{-2}{*}{\textcolor{TGblue}{$\diagdown$} \hspace{-10pt}\textcolor{WTred}{$\diagdown$} $0.936$} &\textcolor{WTred}{$\blacksquare$} $0.101$ & \multirow{-2}{*}{\textcolor{TGblue}{$\diamond$}\textcolor{WTred}{$\diamond$} $0.835$}  \\ \cline{2-8} 
\end{tabular}}
\end{center}
\end{table}

\begin{table}[ht]
\caption{Decrease in retinal thickness in units of \si{\micro\meter} per week. \textcolor{TGblue}{$\blacksquare$} indicates the values for the transgenic mouse and \textcolor{WTred}{$\blacksquare$} indicates the values for the wildtype mouse.} 
\label{slopes}
\begin{center} 
\renewcommand{\arraystretch}{1.5}
\begin{tabular}{lcccc}
 & \multicolumn{1}{l}{} & \multicolumn{3}{c}{Position} \\ \cline{3-5} 
 & \multicolumn{1}{l|}{} & \multicolumn{1}{c|}{Total} & \multicolumn{1}{c|}{Superior} & \multicolumn{1}{c|}{Inferior} \\ \cline{2-5} 
\multicolumn{1}{l|}{\multirow{6}{*}{\rotatebox{90}{Retinal Thickness}}} & \multicolumn{1}{c|}{\multirow{2}{*}{Total}} & \multicolumn{1}{c|}{\textcolor{TGblue}{$\blacksquare$} 0.37} & \multicolumn{1}{c|}{\textcolor{TGblue}{$\blacksquare$} 0.40} & \multicolumn{1}{c|}{\textcolor{TGblue}{$\blacksquare$} 0.35} \\
\multicolumn{1}{l|}{} & \multicolumn{1}{c|}{} & \multicolumn{1}{c|}{\textcolor{WTred}{$\blacksquare$} 0.26} & \multicolumn{1}{c|}{\textcolor{WTred}{$\blacksquare$} 0.27} & \multicolumn{1}{c|}{\textcolor{WTred}{$\blacksquare$} 0.24} \\ \cline{2-5} 
\multicolumn{1}{l|}{} & \multicolumn{1}{c|}{\multirow{2}{*}{Outer}} & \multicolumn{1}{c|}{\textcolor{TGblue}{$\blacksquare$} 0.24} & \multicolumn{1}{c|}{\textcolor{TGblue}{$\blacksquare$} 0.26} & \multicolumn{1}{c|}{\textcolor{TGblue}{$\blacksquare$} 0.23} \\
\multicolumn{1}{l|}{} & \multicolumn{1}{c|}{} & \multicolumn{1}{c|}{\textcolor{WTred}{$\blacksquare$} 0.16} & \multicolumn{1}{c|}{\textcolor{WTred}{$\blacksquare$} 0.14} & \multicolumn{1}{c|}{\textcolor{WTred}{$\blacksquare$} 0.17} \\ \cline{2-5} 
\multicolumn{1}{l|}{} & \multicolumn{1}{c|}{\multirow{2}{*}{Inner}} & \multicolumn{1}{c|}{\textcolor{TGblue}{$\blacksquare$} 0.13} & \multicolumn{1}{c|}{\textcolor{TGblue}{$\blacksquare$} 0.14} & \multicolumn{1}{c|}{\textcolor{TGblue}{$\blacksquare$} 0.12} \\
\multicolumn{1}{l|}{} & \multicolumn{1}{c|}{} & \multicolumn{1}{c|}{\textcolor{WTred}{$\blacksquare$} 0.10} & \multicolumn{1}{c|}{\textcolor{WTred}{$\blacksquare$} 0.13} & \multicolumn{1}{c|}{\textcolor{WTred}{$\blacksquare$} 0.07} \\ \cline{2-5} 
\end{tabular}
\end{center} 
\end{table}

\subsection{OCT angiography}

\begin{figure}
\begin{center}
\begin{tabular}{c}
\includegraphics[height=15.2cm]{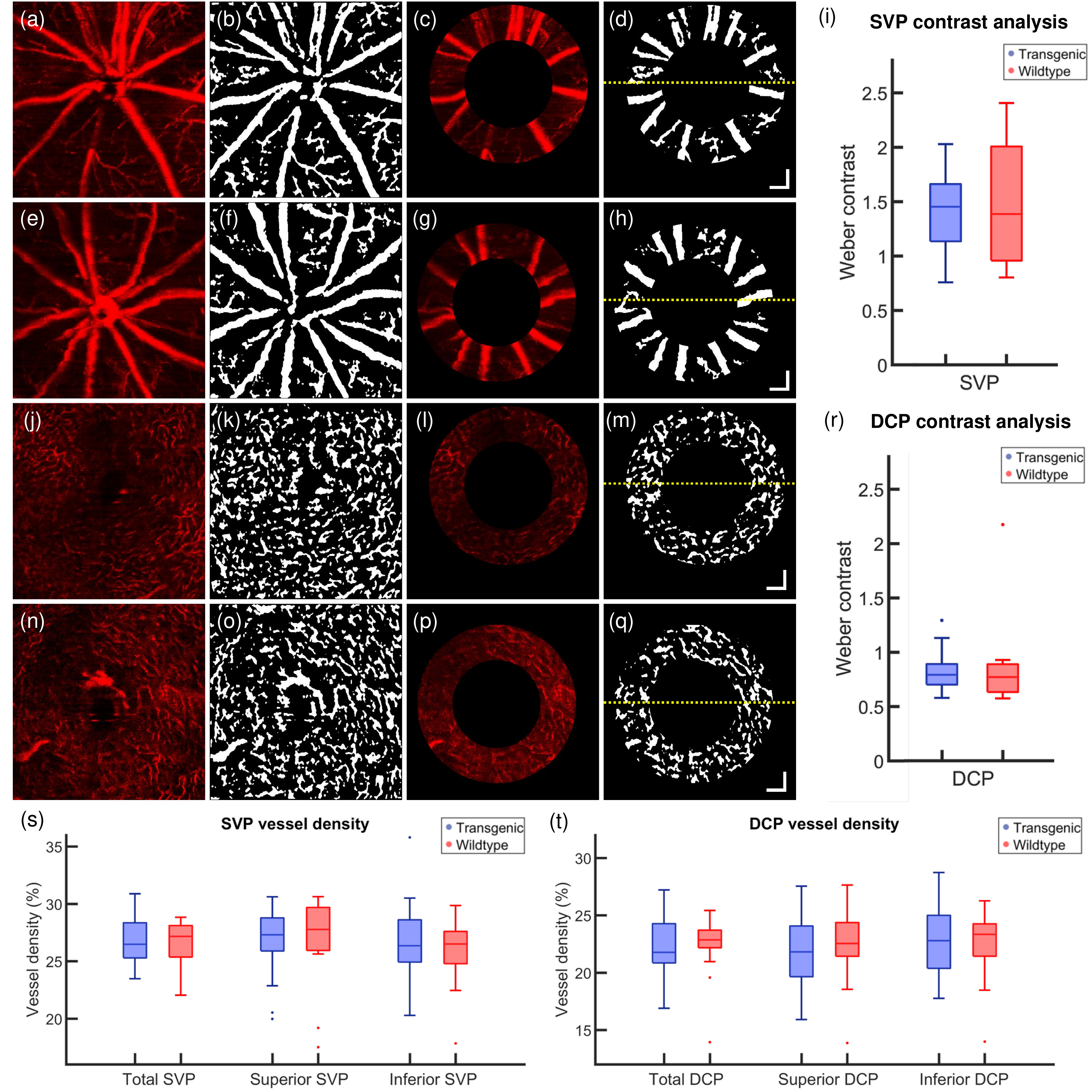}
\end{tabular}
\end{center}
\caption 
{ \label{fig6}
a-h) Example of optical coherence tomography angiography (OCTA) analysis of the superficial vascular plexus (SVP) of a transgenic mouse (a-d) and a wildtype control (e-h). a,e) En-face OCTA depth projection through the SVP. b,f) Binary representation of the SVP with white pixels corresponding to blood vessels. c,g) Annulus around the optic nerve head (ONH) as provided by the intensity-based contrast data. d,h) Binarized annulus, where the yellow dashed line corresponds to the boundary between the superior retina (above) and the inferior retina (below). i) Weber contrast comparing the intensity of the angiogram signal of the blood vessels to the intensity of the background in the SVP. j-q) Example of OCTA analysis of the deep capillary plexus (DCP) of a transgenic mouse (j-m) and a wildtype control (n-q). j,n) En-face OCTA depth projection through the DCP. k,o) Binary representation of the DCP with white pixels corresponding to blood vessels. l,p) Annulus around the ONH as provided by the intensity-based contrast data. m,q) Binarized annulus, where the yellow dashed line corresponds to the boundary between the superior retina (above) and the inferior retina (below). r) Weber contrast comparing the intensity of the angiogram signal of the blood vessels to the intensity of the background in the DCP. s-t) Vessel density analysis. Total, superior and inferior vessel density calculated for transgenic and wildtype mice in the SVP (s) and the DCP (t). Age (a-d, j-m): 93 weeks, Age (e-h, n-q): 76 weeks. Single points in (i) and (r-t) correspond to data outliers. All scale bars = 100 \si{\micro\meter}.} 
\end{figure}

Following layer segmentation of the SVP and the DCP, the vessel density was quantified in these layers in the total, superior and inferior retina for both transgenic and wildtype mice. Figure \ref{fig6}a shows a typical OCT angiogram from the SVP of a transgenic mouse. Combining its binarized form (Fig. \ref{fig6}b) with the annulus which was defined from the reflectivity data (Fig. \ref{fig6}c), binary representations of the retinal vasculature around the ONH were obtained (Fig. \ref{fig6}d). Figure \ref{fig6}e-h shows the same processing steps for a wildtype mouse. The binarized annulus (Fig. \ref{fig6}d,h) was then used as a mask on the annular angiogram (Fig. \ref{fig6}c,g) in order to calculate the Weber contrast (Fig. \ref{fig6}i). The contrast is similar between transgenic and wildtype mice. Figure \ref{fig6}j-m and Fig. \ref{fig6}n-q show examples of the same analysis pattern for the DCP in the transgenic and wildtype mice, respectively. The Weber contrast in this case (Fig. \ref{fig6}r) is lower for both groups, although the contrast values remain similar between transgenic and wildtype mice.

The vessel density calculations for all retinas in the SVP and the DVP are shown in Fig. \ref{fig6}s and Fig. \ref{fig6}t, respectively. No significant differences were observed between transgenic and wildtype mice for any of the retinal regions.

\subsection{Phase retardation abnormalities}

\begin{figure}
\begin{center}
\begin{tabular}{c}
\includegraphics[height=5.5 cm]{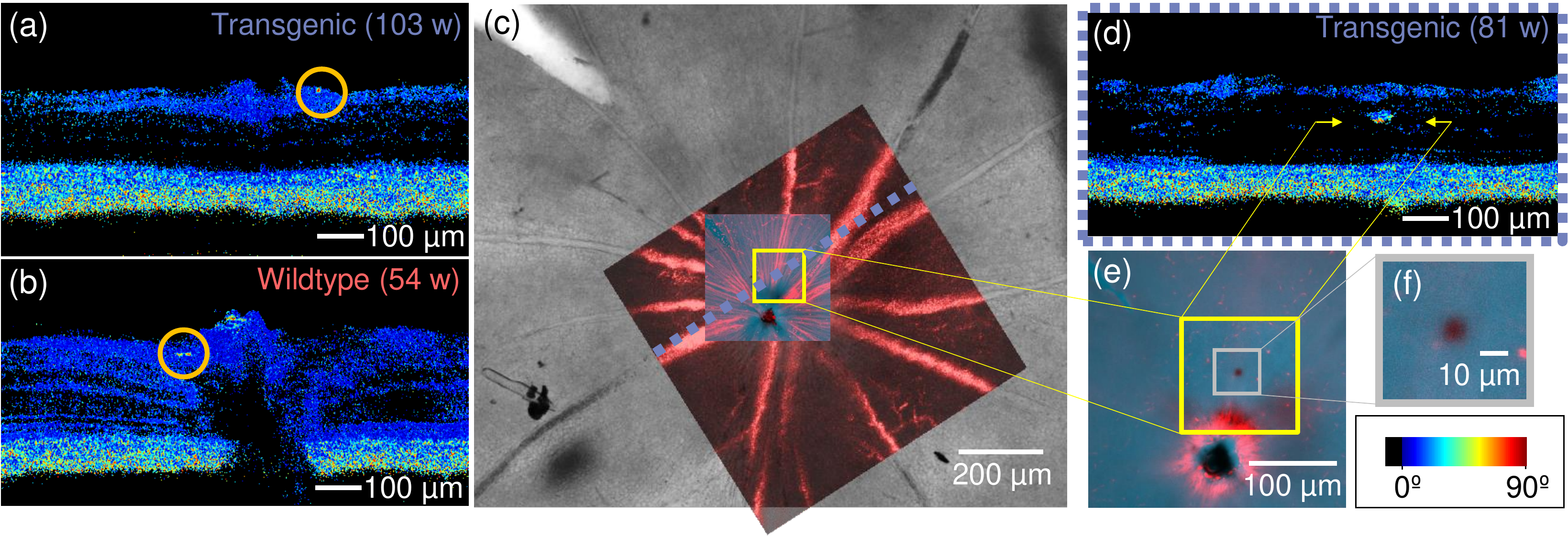}
\end{tabular}
\end{center}
\caption 
{ \label{fig7a} Depolarizing deposits. a) Example of depolarization along a vessel wall (indicated by orange circle). b) Example of depolarization near the optic nerve head (indicated by orange circle). c) Identification of migrated melanin. After wholemounting the retina, the OCT angiography data (in red) was used to correlate the vessels measured in vivo to the overview of the retina provided by the ex vivo preparation (greyscale). d) PS-OCT image showed location of abnormally high phase retardation in the inner nuclear layer (indicated by yellow arrows). e) A high resolution confocal microscopy scan was acquired at the area of interest marked in (c), at the depth position marked by the yellow arrows in (d). A cluster of melanin is revealed at this location, as seen in (f). Scale bar in bottom right applies to (a), (b) and (d).} 
\end{figure}

All individual B-scans of all mouse retinas were screened for phase retardation abnormalities, i.e., depolarizing deposits located outwith the known melanin-containing RPE and ONH regions. Such deposits were found in at least one eye of 22 out of 24 transgenic mice and 11 out of 15 controls, and there were no apparent differences between those deposits observed in transgenic and wildtype groups. Figure \ref{fig7a}a-b shows the two most common forms of phase retardation abnormalities, which appear as small, round depolarizing deposits beside a vessel wall (Fig. \ref{fig7a}a) or beneath the RNFL adjacent to the optic nerve head (Fig. \ref{fig7a}b). No A$\beta$ plaques were identified in any of the retinas where the PS-OCT data was correlated to histology (more details in Section \ref{histo}). However in some cases, melanin migration was found to be the source of the contrast, as demonstrated in Fig. \ref{fig7a}c-f.

\subsection{Double-banded OPL}
\label{opl}
\begin{figure}
\begin{center}
\begin{tabular}{c}
\includegraphics[height=6.25 cm]{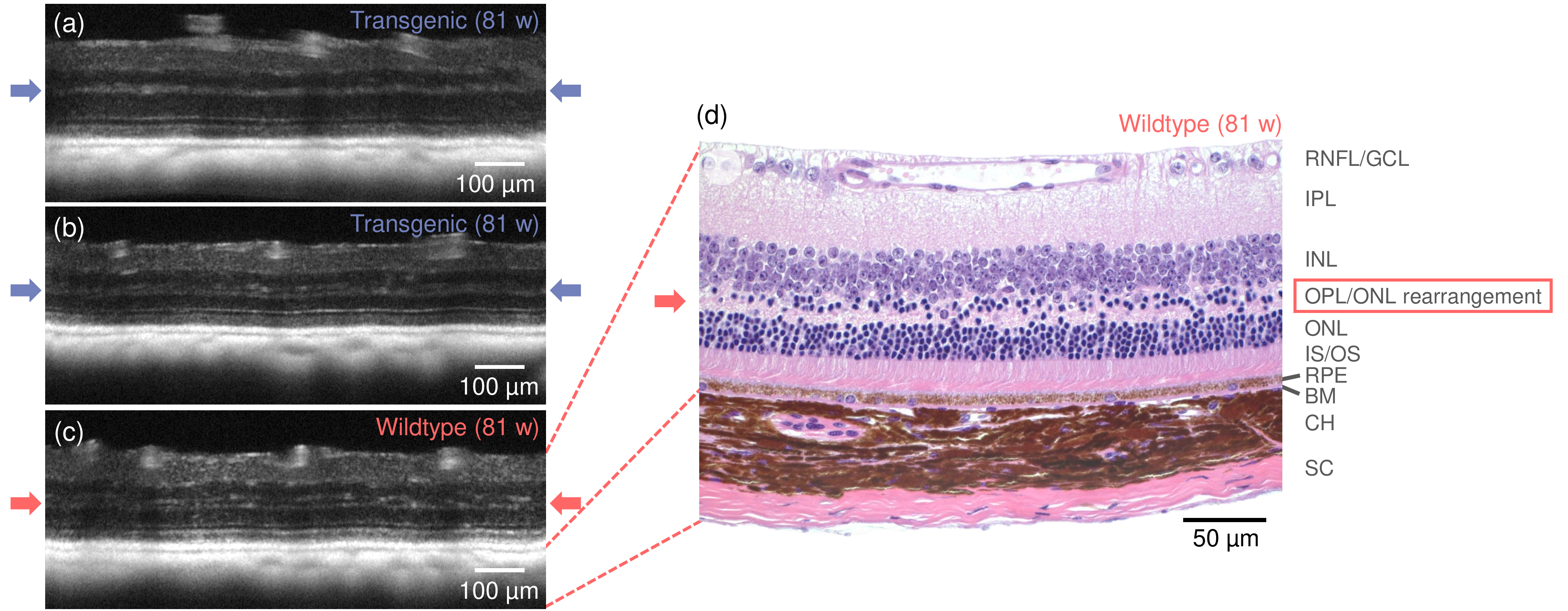}
\end{tabular}
\end{center}
\caption 
{ \label{fig8}
Demonstration of retinal layer abnormalities. The outer plexiform layer (OPL) is indicated with arrows. a) A transgenic mouse retina with a typical appearance - the outer plexiform layer appears as one single hyperreflective band. b) A transgenic mouse retina with the OPL disrupted, appearing as a double-banded hyperreflective layer. This effect was observed in 3/24 transgenic mice. c) A similar double-banding effect was also observed in 3/15 wildtype littermates. d) H\&E-stained histological slice of the same mouse retina as in (c). The structural correlate of the double-banded OCT signal in the OPL region appears to be rearranged proximal outer nuclear layer somata. \textbf{RNFL} Retinal nerve fiber layer. \textbf{GCL} Ganglion cell layer. \textbf{IPL} Inner plexiform layer. \textbf{INL} Inner nuclear layer. \textbf{OPL} Outer plexiform layer. \textbf{ONL} Outer nuclear layer. \textbf{IS/OS} Inner/outer segment junction. \textbf{RPE} Retinal pigment epithelium. \textbf{BM} Bruch's membrane. \textbf{CH} Choroid. \textbf{SC} Sclera. Age (a-d): 81 weeks.} 
\end{figure}

Abnormalities in the structure of the ONL/OPL were found in the reflectivity OCT images in both eyes in a total of 3/24 transgenic mice (age: 54 weeks, 67 weeks and 81 weeks) and 3/15 wildtype control (age: 67 weeks, 81 weeks and 97 weeks). Examples of the appearance of the double-banded OPL can be found in Fig. \ref{fig8}. Figure \ref{fig8}a shows an example of a ``normal'' appearance of a retina observed in a transgenic mouse, where the OPL appears as a single hyperreflective band. In contrast, the hyperreflective OPL appears to split into two in Fig. \ref{fig8}b. Similar double bands of hyperreflective OPL signal were observed in the 3 wildtype mice as shown in  Fig. \ref{fig8}c. To evaluate potential structural bases underlying the atypical retinal layer contrast, exemplary mouse retinas depicting OPL double-banding in the OCT exam were embedded in paraffin, sectioned, and stained with hematoxylin and eosin. Microscopical examination revealed that the double-banding of the OPL precisely correlated witha rearrangement of proximal ONL somata towards the outer border of the INL (Fig. \ref{fig8}d).

\subsection{Retinal histology}
\label{histo}
\subsubsection{Typical observations}

To confirm human A$\beta$ in the retinas of APP/PS1 transgenic mice, indirect immunofluorescent staining of retinal wholemounts was performed with a mouse monoclonal antibody directed against amino acids 1-17 of human A$\beta$ (clone DE2B4). The marker identifies intracellular A$\beta^{83}$ and safely detects extracellular A$\beta$ without cross-reacting with APP$^{83}$. This is evidenced by the clear labeling of A$\beta$ plaques only in brain sections from APP/PS1 transgenic animals used as a positive control.

\begin{figure}
\begin{center}
\begin{tabular}{c}
\includegraphics[height=16.7 cm]{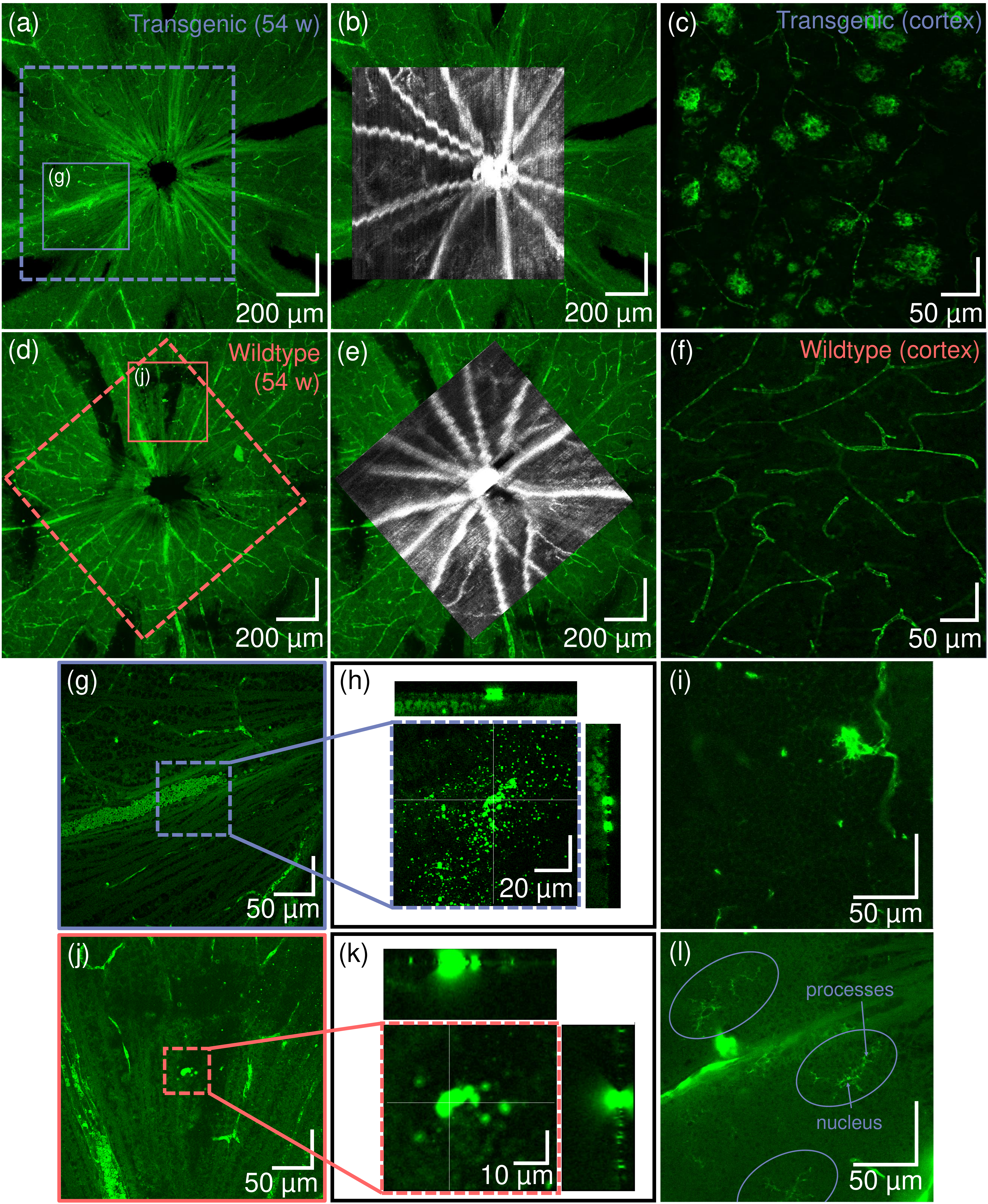}
\end{tabular}
\end{center}
\caption 
{ \label{fig7b}
Representative depictions of the ex vivo retina following fluorescent staining against amyloid beta (A$\beta$). a) Retinal wholemount of a transgenic mouse and b) its correlation to in vivo OCT data. c) The positive control (the cortex of the trangenic mouse) shows fluorescent labeling of A$\beta$ plaques and capillaries. d) Retinal wholemount of a wildtype mouse and e) its correlation to in vivo OCT data. f) In the cortex of the wildtype mouse (negative control), only capillaries are labeled. g-l) Some typical observations seen throughout transgenic and wildtype mouse retinas. g) When zooming in to the surface of the retina in the location indicated in the dashed box (h), some scattered bright spots appear. The orthogonal views (positions indicated by the white cross-hairs), however, show that these lie only on the surface of the retina. i) Fluorescent signal positioned at a capillary junction. j-k) Similar to (g-h), single, larger accumulations of fluorescent tracer find themselves at the interface of vitreous and retina, but not within the retina. l) Microglia, indicated by ovals, are identifiable by their dendritic processes and are also found throughout the retina. This image was acquired within the ganglion cell layer.} 
\end{figure}

Following the donkey anti-mouse secondary antibody staining protocol outlined in Section \ref{retina}, it was expected that in addition to A$\beta$, sources of endogenous IgG present in the retina would bind to the secondary anti-mouse antibody and be highlighted. Figure \ref{fig7b}a shows an example of a retinal wholemount of a transgenic mouse with the peripapillary blood vessels distinctly labeled due to abundant endogenous mouse IgG present in the serum. Figure \ref{fig7b}b  demonstrates the precise fit of the ex vivo vessel pattern with the in vivo OCTA image. The wavy appearance of some of the vessels is a result of motion artefacts caused by breathing during the measurement. From the positive control of the cortex of the transgenic mouse (Fig. \ref{fig7b}c), it can already be observed that signal of a similar intensity to the roundish A$\beta$ plaques also comes from the capillary network. Figure \ref{fig7b}d-f shows the equivalent images for an example of a wildtype control mouse. In the cortex of the wildtype mice (used as a negative control, Fig. \ref{fig7b}f), only the capillary network showed fluorescent labeling. 

Outwith the blood vessels and capillary network, other sources of fluorescent signal were found in both transgenic and control mice. Examples of such features are shown in (Fig. \ref{fig7b}g-l). In Fig. \ref{fig7b}a, a blood vessel (indicated by the solid box) appears to be intensely fluorescing. However, by analysing a series of confocal optical sections (z-stacks) throughout this region (Fig. \ref{fig7b}g-h), it became clear that this signal derived from aggregates of secondary antibody-fluorochrome conjugate artefactually adhering to the surface of the retinal ganglion cell layer. The signal did not derive from either neuronal or non-neuronal structures within the retina it as the overview at 10$\times$ magnification made it appear. Similar observations were made in the wildtype retinas too (Fig. \ref{fig7b}j-k). Structures signalling intensely from within the retina included areas where branches of retinal capillaries appeared to get close together, reminiscent of micro-aneurysms (Fig. \ref{fig7b}i), and microglia (Fig. \ref{fig7b}l), identifiable by the dendritic morphology of their processes. Any source of fluorescent signal which did not fall under one of these categories was then considered a candidate for A$\beta$.

\subsubsection{Potential deposits of retinal amyloid beta}

\begin{figure}
\begin{center}
\begin{tabular}{c}
\includegraphics[height=11.7 cm]{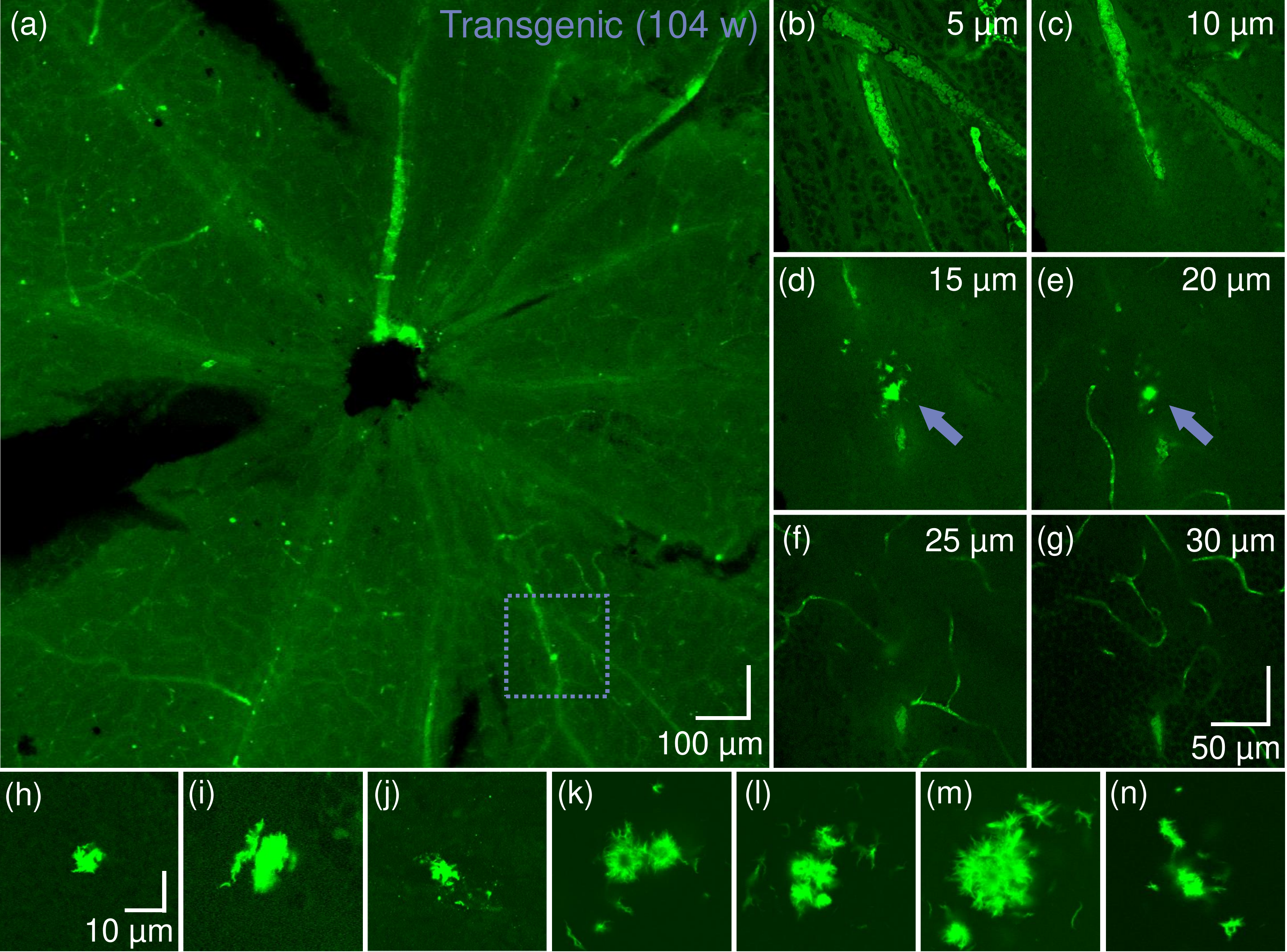}
\end{tabular}
\end{center}
\caption 
{ \label{fig7c}
Candidates for fibrillary amyloid beta (A$\beta$) detected in the retina of one mouse, as identified by confocal microscopy. a) Overview of the ex vivo retina (left eye) acquired with a $10\times$ magnification objective lens. b-g) En-face planes at 5 \si{\micro\meter} intervals at the position identified by the dashed box in (a), where the zero-position is at the interface of vitreous and ganglion cell layer. The fluorescent abnormality, i.e., the A$\beta$ candidate, is indicated by the arrow in (d) and (e). Scale bar in (g) is valid for (b-g). Images were acquired with a $40\times$ magnification objective lens. h-n) Seven further A$\beta$ candidates were identified in the retina of the right eye of the same mouse (all acquired with a $40\times$ magnification objective lens). All structures were detected less than 40 \si{\micro\meter} from the surface of the retina, i.e., between the retinal nerve fiber layer and the inner plexiform layer. Scale bar in (h) is valid for (h-n).} 
\end{figure}

Of the 17 mice which underwent retinal histology, only one mouse (transgenic, age: 104 weeks) displayed fluorescent signals in the retina which could be attributed to extracellular A$\beta$. In this mouse, there was one such area of interest in the left eye, and seven in the right eye. Images of all A$\beta$ plaque candidates can be found in Fig. \ref{fig7c}. Figure \ref{fig7c}a shows an overview of the left retina. While many bright spots were observed, only the one indicated by the dashed box did not exhibit the characteristics of what was shown in Fig. \ref{fig7b}. A z-stack through the retina confirmed that this plaque candidate sat approximately 15 \si{\micro\meter} below the surface of the retina, extending into the anterior IPL. Examples of z-planes can be found in Figure \ref{fig7c}b-g. A similar analysis of the right retina provided images of the further seven plaque candidates, which can be seen in Fig. \ref{fig7c}h-n. The locations of three of the eight plaque candidates were also covered in the field of view of the in vivo OCT measurements, however no abnormalities were found in these locations in the OCT data, using any mode of contrast.

\subsection{Cortical amyloid beta plaque load}

\begin{figure}
\begin{center}
\begin{tabular}{c}
\includegraphics[height=8 cm]{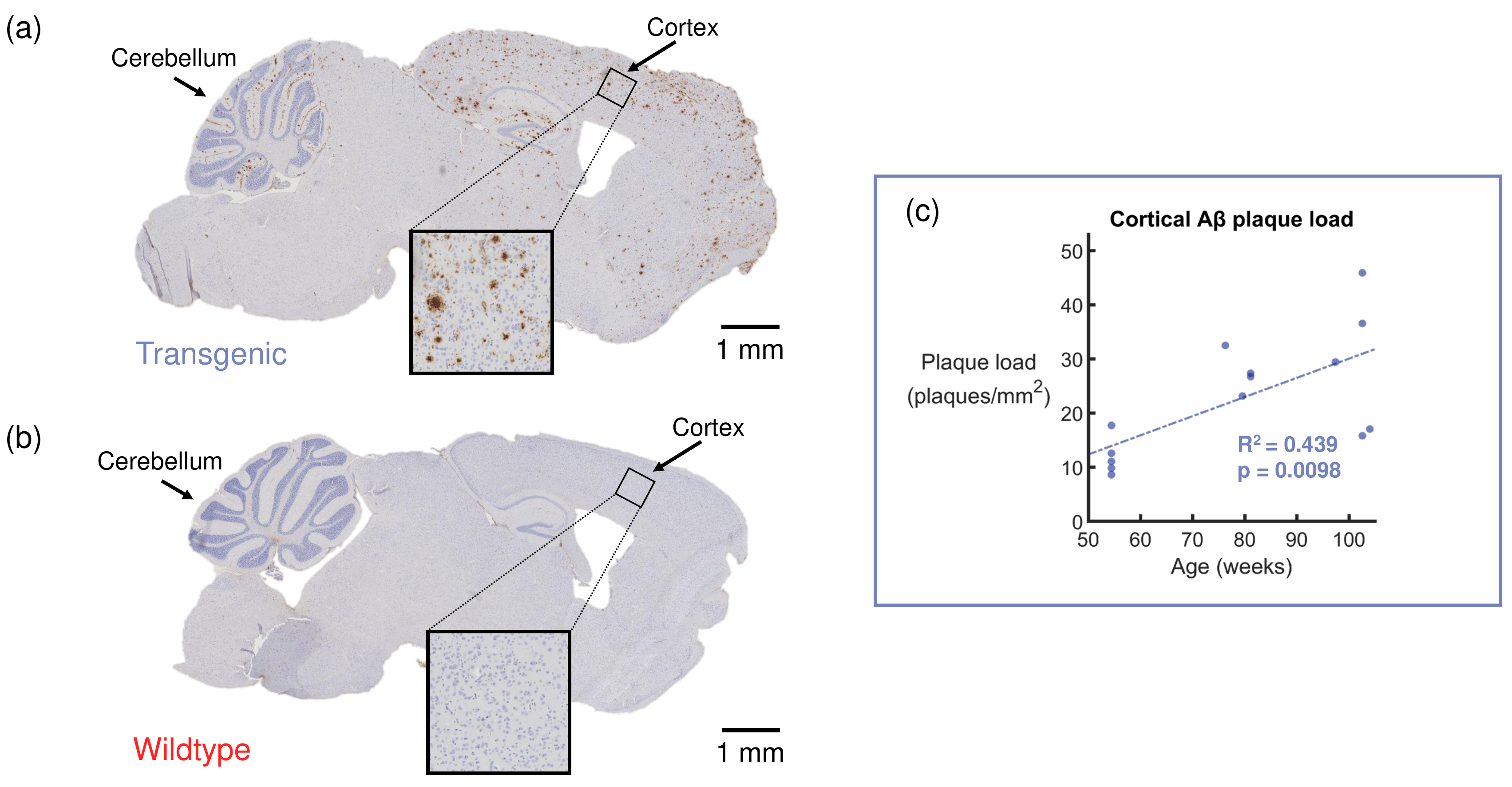}
\end{tabular}
\end{center}
\caption 
{ \label{fig9}
Quantification of the plaques per $\si{\milli\meter\squared}$ in the cortex. a) Histological slice of a transgenic mouse brain, immunohistochemically stained against amyloid-beta (A$\beta$). A$\beta$ plaques appear as brown deposits. Age: 103 weeks. b) Following the same staining protocol, the wildtype littermates do not show any A$\beta$ plaques in the cortex. Age: 103 weeks. c) Count of plaques per $\si{\milli\meter\squared}$ for a subset of 14 transgenic mice (the subset of 7 controls showed no plaques). Linear regression analysis showed a statistically significant trend of an increasing plaque load with age ($R^2 = 0.439$, $p = 0.0098$).} 
\end{figure}

In a subset of the mice (14 transgenic mice and 7 wildtype littermates), histological slices of the brain were prepared and immunohistochemically stained against A$\beta$. Figure \ref{fig9}a shows an example of the staining results for a transgenic mouse, while Fig. \ref{fig9}b shows an example of a similar region in the control brain. In the transgenic mice, brown plaques were identifiable throughout the entire cortex. The mouse depicted in Fig. \ref{fig9}a was 103 weeks old at the time of sacrifice, and A$\beta$ plaques were also visible in abundance in the hippocampal formation and the cerebellum, as well as in other areas of the brain. To the contrary, no plaques were observed in any of the brain regions in the 7 examined wildtype mice (Fig. \ref{fig9}b).

For the 14 transgenic mice, the plaque load was plotted as a function of age. This plot can be found in Fig. \ref{fig9}c. Linear regression analysis revealed an $R^2$ value of 0.439 and a p-value of 0.0098. The gradient of the slope indicated that the plaque load increases by 0.354 plaques per \si{\milli\meter\squared} per week over the investigated age range. Such a result demonstrates that A$\beta$ plaque load increases with age in the transgenic mouse brain, and that the trend is statistically significant. 

\section{Discussion}

Since the role of the retina in Alzheimer's disease is still widely disputed, the aim of this work was to provide a comprehensive overview of what can be observed in the retina of an APP/PS1 mouse model using multi-contrast OCT, and to compare this to histological results. A combination of reflectivity images, PS-OCT images and OCTA were used to investigate the structure and function of the retina. From the in vivo data alone, several retinal abnormalities were successfully identified in this model, however there were no statistically significant differences between the transgenic and wildtype groups. This suggests that the hyperreflective foci, retinal thickness changes, phase retardation abnormalities and structural differences which were measured with OCT are either strain-related or age-related, rather than being due to the genetic mutation itself.

The total retinal thickness was found to significantly decrease with age in the superior and inferior halves as well as in the whole annulus around the optic nerve head. No difference was seen, however, between the transgenic and wildtype groups. These results provide an in vivo validation for that which was previously observed ex vivo by Perez et al. \cite{Perez2009}. In future studies, the inner and outer retina could be subdivided further to quantify individual layer thickness. Since RNFL thinning occurs in AD patients \cite{den2017retinal,chan2019spectral}, it may also be interesting to quantify the RNFL thickness alone in this mouse model. However, quantifying the RNFL thickness in mice using OCT is difficult, as the peripapillary thickness of the healthy RNFL is only approximately 20 \si{\micro\meter} \cite{dysli2015quantitative}, and is interrupted by blood vessels and ganglion cells. Automatic segmentation of the mouse RNFL is therefore challenging, as it can not always be distinguished from the IPL in the OCT images. Owing to its fibrous structure, the RNFL is also birefringent and since it is much thicker in the human retina, the effect of the birefringence is stronger. PS-OCT has already been proposed as a tool for imaging the RNFL in glaucoma \cite{cense2002vivo,zotter2013measuring}, so RNFL measurements in AD patients may be a promising future application for PS-OCT.

The phase retardation analysis performed in this study mainly identified depolarization deposits associated with vessel walls and melanin pigments. No depolarizing deposits were identified in locations which corresponded to candidates for retinal A$\beta$, although locations of melanin migration which were observed in this study mirror results which were documented in the human in age-related macular degeneration \cite{miura2017evaluation}. Future experiments on this topic could also consider HRF in the intensity data, but this becomes challenging in layers which contain vasculature, as the edges of vessels can also appear as hyperscattering features. Previous PS-OCT studies identifying migration of melanin pigments have typically used the metric of degree-of-polarization-uniformity (DOPU) to quantify the polarization scrambling, or depolarization, of the structure in question \cite{gotzinger2008retinal,augustin2016multi,miura2017evaluation}. In order to calculate DOPU, a sliding window must be applied within the images, sacrificing the overall resolution. In this study, as the depolarising structures were small, it was decided to use the phase retardation images alone, as they provided higher image resolution.

The OCTA analysis revealed no significant differences in the vessel density between transgenic and wildtype groups in either the SVP or the DCP. This is contrary to what has been observed in human vessel microangiography studies \cite{zabel2019comparison,yoon2019retinal}. However, in the previous human studies, the area of interest usually targets the fovea. Since mice do not have a fovea, there is not an equivalent measurement in the mouse retina. The modified Weber contrast was similar between transgenic and control groups; the wider standard deviation for the wildtype can be attributed to the smaller sample size. The Weber contrast in the DCP was lower than in the SVP for both groups, which is to be expected as the blood flow in the DCP is lower than in the SVP and therefore the angiogram signal is weaker due to the more random orientation of the red blood cells. Angiograms with poor signal had already been excluded from the analysis as the imaging of older retinas using optical methods can prove challenging. Older mice generally develop cataracts \cite{wolf2000normal}, which results in a lower transmission of light through the lens and therefore a poorer image quality. The shorter the wavelength which is used for OCT, the more attenuating a cataract becomes. Therein lies a trade-off between the axial resolution of the OCT images (the shorter the wavelength, the higher the resolution) and the maximum SNR which can be achieved in the retinal images. Care must also be taken that eye drops are rigorously applied throughout the experiment to ensure a cataract does not form while the animal is under anesthesia \cite{baumann2018polarization}.

The disorganisation of the OPL/ONL structure observed in three transgenic mice and three wildtype mice was not expected from current literature regarding this mouse model. Previous studies in both the mouse \cite{chang2006nob2} and the human \cite{vincent2012outer} have attributed similar OPL/ONL splitting to mutations in the CACNA1F gene encoding for the L-type calcium channel Ca$_\text{v}$1.4 which is also expressed in the outer nuclear layer of the mouse retina. Without the Ca$_\text{v}$1.4 calcium channel, photoreceptor synapses are lost, and the dendritic sprouting which occurs in the photoreceptor layer (in the second-order neurons) is abnormal \cite{mansergh2005mutation}. Whether this gene is defect in this particular APP/PS1 mouse lineage is a topic which must be explored further.

Regarding the analysis of the immunolabeled wholemounted retinas, strong fluorescent signal appears to derive from a range of sources. Examples of fluorescent signal caused by aggregates of the secondary antibody which adhere non-specifically to sticky remnants of the vitreous on the surface of the the ganglion cell layer of the retinal wholemounts can be seen in Fig. \ref{fig7b}g-h and Fig. \ref{fig7b}j-k. With the employed immunostaining protocol, non-A$\beta$ specific signal also derives from binding of the anti-mouse secondary antibody to endogenous IgGs present within, e.g., serum and microglia. This makes it difficult to unequivocally assign biochemical A$\beta$ specificity to highlighted structures. Therefore, in order to better delineate potential deposits of intra- or extracellular A$\beta$, the morphology of the fluorescent signal was also considered. In this study, both retinas of only one transgenic mouse were highlighted as containing candidates for retinal A$\beta$ deposits, displaying intense fluorescence signal and also presenting with a fibril-accumulation-like structure. Of note, A$\beta$ immunopositive assemblies with similar and distinct fibrillary appearance and a size in the few \si{\micro\meter} range have been described in human AD retinas \cite{koronyo2017retinal}. 

Previous studies have indicated that A$\beta$ accumulations in the brain are visible with PS-OCT and regular intensity-based OCT \cite{Baumann2017,Lichtenegger2018,gesperger2019comparison}, however this study was not able to recreate these findings in the retina. This could be due to the difference in size of the A$\beta$ plaques - the plaque candidates which are proposed in Fig. \ref{fig7c} are much smaller than those in the brain (example in Fig. \ref{fig9}a). It has also been previously concluded that not all plaques are visible by either contrast modality \cite{gesperger2019comparison}. Hence negative OCT findings do not rule out the presence of retinal A$\beta$ plaques. Even if plaques could be seen in the retina with OCT, it would be difficult to distinguish them from the HRF and the depolarizing deposits which are already present in these retinas as observed in Fig. \ref{fig4} and Fig. \ref{fig7a}. In our immunofluorescence protocol, our ``positive control'' is the brain, and not the retina, as no retinal ``positive control'' sample exists. Despite our best efforts to mimic retinal wholemount conditions in the control samples, the tissues are simply not the same and therefore it cannot be ruled out that the immunostaining protocol is less optimal for the retina than it is for the brain. However, the candidates for retinal A$\beta$ identified in Fig. \ref{fig7c} would provide an argument that the protocol is indeed suitable.

Given the lack of differences between transgenic and control groups in the in vivo OCT data, and the fact that A$\beta$ could only be identified ex vivo in 1 out of 11 transgenic animals, the suitability of this APP/PS1 mouse as a model of the human must be called into question where the retina is concerned. As with any other mouse model of AD, it only models some aspects of the disease and not others, and each mouse model will experience different age-related and strain-related changes in addition to anything caused by gene mutation. This study finds itself among the conflicting reports regarding the presence of A$\beta$ in the retina \cite{ong2018controversies}. Our results indicate that extracellular A$\beta$ may be found in the retina of this mouse model, although not in all, or even most, samples. A much larger study would need to be conducted in order to statistically determine the likelihood of identifying A$\beta$ plaques in the retina of this mouse model. A topic of future exploration could be a comparison of retinal observations in different APP/PS1 mouse models, also adding a quantification of microglia to the retinal analysis \cite{himmel2016beta}.

The cortical A$\beta$ plaque load, evaluated alongside the retinal data, showed a statistically significant increase with age in the transgenic mice. Despite not being able to correlate this to any retinal changes, this study has documented the typical observations which could be expected to be found in the retina of this mouse model, both in vivo and ex vivo. The 1:1 mapping of the OCT data to the retinal histology was crucial for this experiment, allowing a detailed quantitative and qualitative analysis of structural and functional features of this APP/PS1 mouse model. Tri-fold OCT imaging contrast coupled with retinal wholemounts is therefore a promising method for the analysis of animal models of many retinal diseases.

\section{Conclusion}

While the cortical immunohistochemical staining revealed clear, marked differences in A$\beta$ plaque load between APP/PS1 transgenic mice and their wildtype littermates, a similar difference was not observed in the retina. Candidates for retinal A$\beta$ were only identified in 1 out of 17 transgenic mice. Multi-contrast OCT did, however, reveal retinal abnormalities in these mice, including deposits of migrated melanin and a double-banding of the ONL. Owing to the occurrences in both transgenic and control mice, it is likely that these are strain-dependent and not due to the genetic mutation itself. Nevertheless, the combination of multi-contrast OCT with 1:1 mapping of retinal histology allowed for a thorough documentation of what one would expect to see in this APP/PS1 mouse model of AD.  

\appendix    

\subsection*{Disclosures}
The authors have no relevant financial interests in the manuscript and no other potential conflicts of interest to disclose. Part of this work has been presented at the Association for Research in Vision and Ophthalmology (ARVO) Annual Meeting in Vancouver, Canada, and the published abstract by Baumann et al. can be found in Ref.  \cite{baumann2019investigating} . 

\acknowledgments 
Financial support from the European Research Council (ERC StG 640396 OPTIMALZ) and the Austrian Science Fund (FWF, P25823‐B24) is gratefully acknowledged. The authors would like to thank the team at the Neuropathology department at the Medical University of Vienna for providing assistance and advice regarding histology. Sincere thanks are also extended to the staff at the Division of Biomedical Research at the Medical University of Vienna for the animal care. Finally, the authors would like to thank Christoph K. Hitzenberger for his continued support throughout the duration of this project.


\bibliography{report}   

\begin{thebibliography}{10}

\bibitem{Jack2009}
C.~R. Jack, V.~J. Lowe, S.~D. Weigand, {\em et~al.}, ``Serial {PIB} and {MRI}
  in normal, mild cognitive impairment and {A}lzheimer's disease: implications
  for sequence of pathological events in {A}lzheimer's disease,'' {\em Brain}
  {\bf 132}, 1355--1365  (2009).

\bibitem{alzFacts}
{A}lzheimer’s Association, ``2019 {A}lzheimer's disease facts and figures,''
  {\em Alzheimers Dement.} {\bf 15}, 321--387  (2019).

\bibitem{funato1998quantitation}
H.~Funato, M.~Yoshimura, K.~Kusui, {\em et~al.}, ``Quantitation of amyloid
  beta-protein (a beta) in the cortex during aging and in {A}lzheimer's
  disease.,'' {\em Am. J. Pathol.} {\bf 152}(6), 1633  (1998).

\bibitem{Hurtado-Puerto2018}
A.~M. Hurtado-Puerto, C.~Russo, and F.~Fregni, {\em {A}lzheimer's Disease},
  297--338.
\newblock Springer New York, New York, NY  (2018).

\bibitem{dudeffant2017contrast}
C.~Dudeffant, M.~Vandesquille, K.~Herbert, {\em et~al.}, ``Contrast-enhanced
  {MR} microscopy of amyloid plaques in five mouse models of amyloidosis and in
  human {A}lzheimer’s disease brains,'' {\em Sci. Rep.} {\bf 7}(1), 4955
  (2017).

\bibitem{golzan2017retinal}
S.~M. Golzan, K.~Goozee, D.~Georgevsky, {\em et~al.}, ``Retinal vascular and
  structural changes are associated with amyloid burden in the elderly:
  ophthalmic biomarkers of preclinical {A}lzheimer’s disease,'' {\em
  {A}lzheimers Res. Ther.} {\bf 9}(1), 13  (2017).

\bibitem{jentsch2015retinal}
S.~Jentsch, D.~Schweitzer, K.-U. Schmidtke, {\em et~al.}, ``Retinal
  fluorescence lifetime imaging ophthalmoscopy measures depend on the severity
  of {A}lzheimer's disease,'' {\em Acta Ophthalmol.} {\bf 93}(4), e241--e247
  (2015).

\bibitem{ong2018controversies}
S.~S. Ong, P.~M. Doraiswamy, and E.~M. Lad, ``Controversies and future
  directions of ocular biomarkers in {A}lzheimer disease,'' {\em JAMA Neurol.}
  {\bf 75}(6), 650--651  (2018).

\bibitem{koronyo2011identification}
M.~Koronyo-Hamaoui, Y.~Koronyo, A.~V. Ljubimov, {\em et~al.}, ``Identification
  of amyloid plaques in retinas from {A}lzheimer's patients and noninvasive in
  vivo optical imaging of retinal plaques in a mouse model,'' {\em Neuroimage}
  {\bf 54}, S204--S217  (2011).

\bibitem{tsai2014ocular}
Y.~Tsai, B.~Lu, A.~V. Ljubimov, {\em et~al.}, ``Ocular changes in {TgF344-AD}
  rat model of {A}lzheimer's disease,'' {\em Invest. Ophthalmol. Vis. Sci.}
  {\bf 55}(1), 523--534  (2014).

\bibitem{la2016melanopsin}
C.~La~Morgia, F.~N. Ross-Cisneros, Y.~Koronyo, {\em et~al.}, ``Melanopsin
  retinal ganglion cell loss in {A}lzheimer disease,'' {\em Ann. Neurol.} {\bf
  79}(1), 90--109  (2016).

\bibitem{koronyo2017retinal}
Y.~Koronyo, D.~Biggs, E.~Barron, {\em et~al.}, ``Retinal amyloid pathology and
  proof-of-concept imaging trial in {A}lzheimer’s disease,'' {\em J. Clin.
  Invest. Insight} {\bf 2}(16)  (2017).

\bibitem{blanks1989retinal}
J.~C. Blanks, D.~R. Hinton, A.~A. Sadun, {\em et~al.}, ``Retinal ganglion cell
  degeneration in {A}lzheimer's disease,'' {\em Brain Res.} {\bf 501}(2),
  364--372  (1989).

\bibitem{schon2012long}
C.~Sch{\"o}n, N.~A. Hoffmann, S.~M. Ochs, {\em et~al.}, ``Long-term in vivo
  imaging of fibrillar tau in the retina of {P301S} transgenic mice,'' {\em
  PLOS One} {\bf 7}(12), e53547  (2012).

\bibitem{ho2014beta}
C.-Y. Ho, J.~C. Troncoso, D.~Knox, {\em et~al.}, ``Beta-amyloid, phospho-tau
  and alpha-synuclein deposits similar to those in the brain are not identified
  in the eyes of {A}lzheimer's and {P}arkinson's disease patients,'' {\em Brain
  Pathol.} {\bf 24}(1), 25--32  (2014).

\bibitem{den2018amyloid}
J.~den Haan, T.~H. Morrema, F.~D. Verbraak, {\em et~al.}, ``Amyloid-beta and
  phosphorylated tau in post-mortem {A}lzheimer’s disease retinas,'' {\em
  Acta Neuropathol. Commun.} {\bf 6}(1), 147  (2018).

\bibitem{de2004alzheimer}
J.~C. de~la Torre, ``Is {A}lzheimer's disease a neurodegenerative or a vascular
  disorder? {D}ata, dogma, and dialectics,'' {\em Lancet Neurol.} {\bf 3}(3),
  184--190  (2004).

\bibitem{Cheung2014}
C.~Y. Cheung, Y.~T. Ong, M.~K. Ikram, {\em et~al.}, ``Microvascular network
  alterations in the retina of patients with {A}lzheimer's disease,'' {\em
  Alzheimers Dement.} {\bf 10}, 135--142  (2014).

\bibitem{berisha2007retinal}
F.~Berisha, G.~T. Feke, C.~L. Trempe, {\em et~al.}, ``Retinal abnormalities in
  early {A}lzheimer’s disease,'' {\em Invest. Ophthalmol. Vis. Sci.} {\bf
  48}(5), 2285--2289  (2007).

\bibitem{Feke2015}
G.~T. Feke, B.~T. Hyman, R.~A. Stern, {\em et~al.}, ``Retinal blood flow in
  mild cognitive impairment and {A}lzheimer's disease,'' {\em Alzheimers
  Dement.} {\bf 1}, 144--151  (2015).

\bibitem{Williams2015}
M.~A. Williams, A.~J. McGowan, C.~R. Cardwell, {\em et~al.}, ``Retinal
  microvascular network attenuation in {A}lzheimer's disease,'' {\em Alzheimers
  Dement.} {\bf 1}, 229--235  (2015).

\bibitem{zabel2019comparison}
P.~Zabel, J.~J. Kaluzny, M.~Wilkosc-Debczynska, {\em et~al.}, ``Comparison of
  retinal microvasculature in patients with {A}lzheimer's disease and primary
  open-angle glaucoma by optical coherence tomography angiography,'' {\em
  Invest. Ophthalmol. Vis. Sci.} {\bf 60}(10), 3447--3455  (2019).

\bibitem{yoon2019retinal}
S.~P. Yoon, D.~S. Grewal, A.~C. Thompson, {\em et~al.}, ``Retinal microvascular
  and neurodegenerative changes in {A}lzheimer’s disease and mild cognitive
  impairment compared with control participants,'' {\em Ophthalmol. Retina}
  {\bf 3}, 489--499  (2019).

\bibitem{vandeKreekeARVO}
A.~van~de Kreeke, H.~T. Nguyen, E.~Konijnenberg, {\em et~al.}, ``Optical
  coherence tomography angiography in preclinical {A}lzheimer’s disease,''
  {\em Invest. Ophthalmol. Vis. Sci.} {\bf 60}(9), 5233  (2019).

\bibitem{LaughlinARVO}
A.~Laughlin, J.~Ringman, B.~S. Ashimatey, {\em et~al.}, ``Retinal changes in
  early-onset {A}lzheimer disease,'' {\em Invest. Ophthalmol. Vis. Sci.} {\bf
  60}(9), 4563  (2019).

\bibitem{akiyama1994inflammatory}
H.~Akiyama, ``Inflammatory response in {A}lzheimer's disease,'' {\em Tohoku J.
  Exp. Med.} {\bf 174}(3), 295--303  (1994).

\bibitem{heneka2015neuroinflammation}
M.~T. Heneka, M.~J. Carson, J.~El~Khoury, {\em et~al.}, ``Neuroinflammation in
  {A}lzheimer's disease,'' {\em Lancet Neurol.} {\bf 14}(4), 388--405  (2015).

\bibitem{den2017retinal}
J.~den Haan, F.~D. Verbraak, P.~J. Visser, {\em et~al.}, ``Retinal thickness in
  {A}lzheimer's disease: a systematic review and meta-analysis,'' {\em
  {A}lzheimers Dement.} {\bf 6}, 162--170  (2017).

\bibitem{chan2019spectral}
V.~T. Chan, Z.~Sun, S.~Tang, {\em et~al.}, ``Spectral-domain {OCT} measurements
  in {A}lzheimer’s disease: a systematic review and meta-analysis,'' {\em
  Ophthalmology} {\bf 126}(4), 497--510  (2019).

\bibitem{leung2012retinal}
C.~K.-S. Leung, M.~Yu, R.~N. Weinreb, {\em et~al.}, ``Retinal nerve fiber layer
  imaging with spectral-domain optical coherence tomography: patterns of
  retinal nerve fiber layer progression,'' {\em Ophthalmology} {\bf 119}(9),
  1858--1866  (2012).

\bibitem{inzelberg2004retinal}
R.~Inzelberg, J.~A. Ramirez, P.~Nisipeanu, {\em et~al.}, ``Retinal nerve fiber
  layer thinning in {P}arkinson disease,'' {\em Vision Res.} {\bf 44}(24),
  2793--2797  (2004).

\bibitem{alasil2013analysis}
T.~Alasil, K.~Wang, P.~A. Keane, {\em et~al.}, ``Analysis of normal retinal
  nerve fiber layer thickness by age, sex, and race using spectral domain
  optical coherence tomography,'' {\em J. Glaucoma} {\bf 22}(7), 532--541
  (2013).

\bibitem{hall2012mouse}
A.~M. Hall and E.~D. Roberson, ``Mouse models of {A}lzheimer's disease,'' {\em
  Brain Res. Bull.} {\bf 88}(1), 3--12  (2012).

\bibitem{Jankowsky2001}
J.~L. Jankowsky, H.~H. Slunt, T.~Ratovitski, {\em et~al.}, ``Co-expression of
  multiple transgenes in mouse {CNS}: a comparison of strategies,'' {\em
  Biomol. Eng.} {\bf 17}, 157--165  (2001).

\bibitem{Jankowsky2003}
J.~L. Jankowsky, D.~J. Fadale, J.~Anderson, {\em et~al.}, ``Mutant presenilins
  specifically elevate the levels of the 42 residue $\beta$-amyloid peptide in
  vivo: evidence for augmentation of a 42-specific $\gamma$ secretase,'' {\em
  Hum. Mol. Genet.} {\bf 13}, 159--170  (2003).

\bibitem{Reiserer2007}
R.~S. Reiserer, F.~E. Harrison, D.~C. Syverud, {\em et~al.}, ``Impaired spatial
  learning in the {APPSwe}$+${PSEN}1$\delta$9 bigenic mouse model of
  {A}lzheimer's disease,'' {\em Genes Brain Behav.} {\bf 6}, 54--65  (2007).

\bibitem{yan2009characterizing}
P.~Yan, A.~W. Bero, J.~R. Cirrito, {\em et~al.}, ``Characterizing the
  appearance and growth of amyloid plaques in {APP/PS1} mice,'' {\em J.
  Neurosci.} {\bf 29}(34), 10706--10714  (2009).

\bibitem{ordonez2016abetapp}
L.~Ordonez-Gutierrez, I.~Fernandez-Perez, J.~L. Herrera, {\em et~al.},
  ``{A}$\beta${PP/PS1} transgenic mice show sex differences in the cerebellum
  associated with aging,'' {\em J. Alzheimers Dis.} {\bf 54}(2), 645--656
  (2016).

\bibitem{Dutescu2009}
R.~M. Dutescu, Q.-X. Li, J.~Crowston, {\em et~al.}, ``Amyloid precursor protein
  processing and retinal pathology in mouse models of {A}lzheimer's disease,''
  {\em Graefes Arch. Clin. Exp. Ophthalmol.} {\bf 247}, 1213--1221  (2009).

\bibitem{Shimazawa2008}
M.~Shimazawa, Y.~Inokuchi, T.~Okuno, {\em et~al.}, ``Reduced retinal function
  in amyloid precursor protein-over-expressing transgenic mice via attenuating
  {glutamate-N-methyl-d-aspartate} receptor signaling,'' {\em J. Neurochem.}
  {\bf 107}, 279--290  (2008).

\bibitem{Joly2017}
S.~Joly, S.~Lamoureux, and V.~Pernet, ``Nonamyloidogenic processing of amyloid
  beta precursor protein is associated with retinal function improvement in
  aging male {APP}swe/{PS}1$\delta$e9 mice,'' {\em Neurobiol. Aging} {\bf 53},
  181--191  (2017).

\bibitem{chidlow2017investigations}
G.~Chidlow, J.~P. Wood, J.~Manavis, {\em et~al.}, ``Investigations into retinal
  pathology in the early stages of a mouse model of {A}lzheimer’s disease,''
  {\em J. Alzheimers Dis.} {\bf 56}(2), 655--675  (2017).

\bibitem{Ning2008}
A.~Ning, J.~Cui, E.~To, {\em et~al.}, ``Amyloid-$\beta$ deposits lead to
  retinal degeneration in a mouse model of {A}lzheimer disease,'' {\em Invest.
  Ophthalmol. Vis. Sci.} {\bf 49}, 5136  (2008).

\bibitem{Perez2009}
S.~E. Perez, S.~Lumayag, B.~Kovacs, {\em et~al.}, ``$\beta$-amyloid deposition
  and functional impairment in the retina of the {APPswe}/{PS}1$\delta$e9
  transgenic mouse model of {A}lzheimer's disease,'' {\em Invest. Ophthalmol.
  Vis. Sci.} {\bf 50}, 793  (2009).

\bibitem{Chiquita2019}
S.~Chiquita, A.~C. Rodrigues-Neves, F.~I. Baptista, {\em et~al.}, ``The retina
  as a window or mirror of the brain changes detected in {A}lzheimer’s
  disease: Critical aspects to unravel,'' {\em Mol. Neurobiol.} {\bf 56},
  5416--5435  (2019).

\bibitem{shah2017beta}
T.~Shah, S.~Gupta, P.~Chatterjee, {\em et~al.}, ``Beta-amyloid sequelae in the
  eye: a critical review on its diagnostic significance and clinical relevance
  in {A}lzheimer’s disease,'' {\em Mol. Psychiatr.} {\bf 22}(3), 353  (2017).

\bibitem{huang1991optical}
D.~Huang, E.~A. Swanson, C.~P. Lin, {\em et~al.}, ``Optical coherence
  tomography,'' {\em Science} {\bf 254}(5035), 1178--1181  (1991).

\bibitem{de2015review}
T.~E. De~Carlo, A.~Romano, N.~K. Waheed, {\em et~al.}, ``A review of optical
  coherence tomography angiography ({OCTA}),'' {\em Int. J. Retina Vitreous}
  {\bf 1}(1), 5  (2015).

\bibitem{zhu2017can}
J.~Zhu, C.~W. Merkle, M.~T. Bernucci, {\em et~al.}, ``Can {OCT} angiography be
  made a quantitative blood measurement tool?,'' {\em Appl. Sci.} {\bf 7}(7),
  687  (2017).

\bibitem{chen2017optical}
C.-L. Chen and R.~K. Wang, ``Optical coherence tomography based angiography,''
  {\em Biomed. Opt. Express} {\bf 8}(2), 1056--1082  (2017).

\bibitem{hee1992polarization}
M.~R. Hee, D.~Huang, E.~A. Swanson, {\em et~al.}, ``Polarization-sensitive
  low-coherence reflectometer for birefringence characterization and ranging,''
  {\em J. Opt. Soc. Am. B} {\bf 9}(6), 903--908  (1992).

\bibitem{pircher2011polarization}
M.~Pircher, C.~K. Hitzenberger, and U.~Schmidt-Erfurth, ``Polarization
  sensitive optical coherence tomography in the human eye,'' {\em Prog. Retin.
  Eye Res.} {\bf 30}(6), 431--451  (2011).

\bibitem{de2017polarization}
J.~F. de~Boer, C.~K. Hitzenberger, and Y.~Yasuno, ``Polarization sensitive
  optical coherence tomography--a review,'' {\em Biomed. Opt. Express} {\bf
  8}(3), 1838--1873  (2017).

\bibitem{baumann2017polarization}
B.~Baumann, ``Polarization sensitive optical coherence tomography: A review of
  technology and applications,'' {\em Appl. Sci.} {\bf 7}(5), 474  (2017).

\bibitem{jin2003imaging}
L.-W. Jin, K.~A. Claborn, M.~Kurimoto, {\em et~al.}, ``Imaging linear
  birefringence and dichroism in cerebral amyloid pathologies,'' {\em Proc.
  Natl. Acad. Sci.} {\bf 100}(26), 15294--15298  (2003).

\bibitem{campbell2015polarization}
M.~C. Campbell, D.~De~Vries, L.~Emptage, {\em et~al.}, ``Polarization
  properties of amyloid beta in the retina of the eye as a biomarker of
  {A}lzheimer’s disease,'' in {\em Bio-Optics: Design and Application},
  BM3A--4, Optical Society of America  (2015).

\bibitem{hamel2016polarization}
M.~T. Hamel, L.~Emptage, D.~DeVries, {\em et~al.}, ``Polarization properties of
  amyloid deposits in the retinas of an animal model of {A}lzheimer’s disease
  differ in those with and without cognitive impairment,'' {\em Invest.
  Ophthalmol. Vis. Sci.} {\bf 57}(12), 2216--2216  (2016).

\bibitem{campbell2018amyloid}
M.~C. Campbell, L.~Emptage, R.~Redekop, {\em et~al.}, ``Amyloid deposits imaged
  in postmortem retinas using polarimetry predict the severity of a postmortem
  brain based diagnosis of {A}lzheimer's disease,'' {\em Alzheimers Dement.}
  {\bf 14}(7), P774--P775  (2018).

\bibitem{Baumann2017}
B.~Baumann, A.~Woehrer, G.~Ricken, {\em et~al.}, ``Visualization of neuritic
  plaques in {A}lzheimer's disease by polarization-sensitive optical coherence
  microscopy,'' {\em Sci. Rep.} {\bf 7}, 43477  (2017).

\bibitem{gesperger2019comparison}
J.~Gesperger, A.~Lichtenegger, T.~Roetzer, {\em et~al.}, ``Comparison of
  intensity-and polarization-based contrast in amyloid-beta plaques as observed
  by optical coherence tomography,'' {\em Appl. Sci.} {\bf 9}(10), 2100
  (2019).

\bibitem{bolmont2012label}
T.~Bolmont, A.~Bouwens, C.~Pache, {\em et~al.}, ``Label-free imaging of
  cerebral $\beta$-amyloidosis with extended-focus optical coherence
  microscopy,'' {\em J. Neurosci.} {\bf 32}(42), 14548--14556  (2012).

\bibitem{Marchand2017}
P.~J. Marchand, A.~Bouwens, D.~Szlag, {\em et~al.}, ``Visible spectrum
  extended-focus optical coherence microscopy for label-free sub-cellular
  tomography,'' {\em Biomed. Opt. Express} {\bf 8}, 3343  (2017).

\bibitem{Lichtenegger2018}
A.~Lichtenegger, M.~Muck, P.~Eugui, {\em et~al.}, ``Assessment of pathological
  features in {A}lzheimer's disease brain tissue with a large field-of-view
  visible-light optical coherence microscope,'' {\em Neurophoton.} {\bf 5},
  035002  (2018).

\bibitem{fialova2016polarization}
S.~Fialov{\'a}, M.~Augustin, M.~Gl{\"o}smann, {\em et~al.}, ``Polarization
  properties of single layers in the posterior eyes of mice and rats
  investigated using high resolution polarization sensitive optical coherence
  tomography,'' {\em Biomed. Opt. Express} {\bf 7}(4), 1479--1495  (2016).

\bibitem{harper2018white}
D.~J. Harper, M.~Augustin, A.~Lichtenegger, {\em et~al.}, ``White light
  polarization sensitive optical coherence tomography for sub-micron axial
  resolution and spectroscopic contrast in the murine retina,'' {\em Biomed.
  Opt. Express} {\bf 9}(5), 2115--2129  (2018).

\bibitem{augustin2016multi}
M.~Augustin, S.~Fialov{\'a}, T.~Himmel, {\em et~al.}, ``Multi-functional {OCT}
  enables longitudinal study of retinal changes in a {VLDLR} knockout mouse
  model,'' {\em PLOS One} {\bf 11}(10), e0164419  (2016).

\bibitem{augustin2018segmentation}
M.~Augustin, D.~J. Harper, C.~W. Merkle, {\em et~al.}, ``Segmentation of
  retinal layers in {OCT} images of the mouse eye utilizing polarization
  contrast,'' in {\em Computational Pathology and Ophthalmic Medical Image
  Analysis},  D.~S. et~al., Ed., 310--318, Springer  (2018).

\bibitem{yushkevich2006user}
P.~A. Yushkevich, J.~Piven, H.~C. Hazlett, {\em et~al.}, ``User-guided {3D}
  active contour segmentation of anatomical structures: significantly improved
  efficiency and reliability,'' {\em Neuroimage} {\bf 31}(3), 1116--1128
  (2006).

\bibitem{gotzinger2005high}
E.~G{\"o}tzinger, M.~Pircher, and C.~K. Hitzenberger, ``High speed spectral
  domain polarization sensitive optical coherence tomography of the human
  retina,'' {\em Opt. Express} {\bf 13}(25), 10217--10229  (2005).

\bibitem{zuiderveld1994contrast}
K.~Zuiderveld, ``Contrast limited adaptive histogram equalization,'' in {\em
  Graphics Gems IV},  474--485, Academic Press Professional, Inc.  (1994).

\bibitem{weber}
G.~T. Fechner and W.~M. Wundt, {\em Elemente der Psychophysik}, Breitkopf \&
  Härtel, Leipzig  (1889).

\bibitem{schindelin2012fiji}
J.~Schindelin, I.~Arganda-Carreras, E.~Frise, {\em et~al.}, ``Fiji: an
  open-source platform for biological-image analysis,'' {\em Nat. Methods} {\bf
  9}(7), 676--682  (2012).

\bibitem{loo2012effects}
A.~E.~K. Loo, Y.~T. Wong, R.~Ho, {\em et~al.}, ``Effects of hydrogen peroxide
  on wound healing in mice in relation to oxidative damage,'' {\em PLOS One}
  {\bf 7}(11), e49215  (2012).

\bibitem{dysli2015quantitative}
C.~Dysli, V.~Enzmann, R.~Sznitman, {\em et~al.}, ``Quantitative analysis of
  mouse retinal layers using automated segmentation of spectral domain optical
  coherence tomography images,'' {\em Transl. Vis. Sci. Techn.} {\bf 4}(4), 9
  (2015).

\bibitem{cense2002vivo}
B.~Cense, T.~C. Chen, B.~H. Park, {\em et~al.}, ``In vivo depth-resolved
  birefringence measurements of the human retinal nerve fiber layer by
  polarization-sensitive optical coherence tomography,'' {\em Opt. Lett.} {\bf
  27}(18), 1610--1612  (2002).

\bibitem{zotter2013measuring}
S.~Zotter, M.~Pircher, E.~G{\"o}tzinger, {\em et~al.}, ``Measuring retinal
  nerve fiber layer birefringence, retardation, and thickness using wide-field,
  high-speed polarization sensitive spectral domain {OCT},'' {\em Invest.
  Ophthalmol. Vis. Sci.} {\bf 54}(1), 72--84  (2013).

\bibitem{miura2017evaluation}
M.~Miura, S.~Makita, S.~Sugiyama, {\em et~al.}, ``Evaluation of intraretinal
  migration of retinal pigment epithelial cells in age-related macular
  degeneration using polarimetric imaging,'' {\em Sci. Rep.} {\bf 7}(1), 3150
  (2017).

\bibitem{gotzinger2008retinal}
E.~G{\"o}tzinger, M.~Pircher, W.~Geitzenauer, {\em et~al.}, ``Retinal pigment
  epithelium segmentation by polarization sensitive optical coherence
  tomography,'' {\em Opt. Express} {\bf 16}(21), 16410--16422  (2008).

\bibitem{wolf2000normal}
N.~S. Wolf, Y.~Li, W.~Pendergrass, {\em et~al.}, ``Normal mouse and rat strains
  as models for age-related cataract and the effect of caloric restriction on
  its development,'' {\em Exp. Eye. Res.} {\bf 70}(5), 683--692  (2000).

\bibitem{baumann2018polarization}
B.~Baumann, M.~Augustin, A.~Lichtenegger, {\em et~al.},
  ``Polarization-sensitive optical coherence tomography imaging of the anterior
  mouse eye,'' {\em J. Biomed. Opt.} {\bf 23}(8), 086005  (2018).

\bibitem{chang2006nob2}
B.~Chang, J.~R. Heckenlively, P.~R. Bayley, {\em et~al.}, ``The nob2 mouse, a
  null mutation in {CACNA1F}: anatomical and functional abnormalities in the
  outer retina and their consequences on ganglion cell visual responses,'' {\em
  Visual Neurosci.} {\bf 23}(1), 11--24  (2006).

\bibitem{vincent2012outer}
A.~Vincent and E.~H{\'e}on, ``Outer retinal structural anomaly due to
  frameshift mutation in {CACNA1F} gene,'' {\em Eye} {\bf 26}(9), 1278  (2012).

\bibitem{mansergh2005mutation}
F.~Mansergh, N.~C. Orton, J.~P. Vessey, {\em et~al.}, ``Mutation of the calcium
  channel gene {C}acna1f disrupts calcium signaling, synaptic transmission and
  cellular organization in mouse retina,'' {\em Hum. Molec. Genet.} {\bf
  14}(20), 3035--3046  (2005).

\bibitem{himmel2016beta}
T.~Himmel, S.~Fialova, M.~Augustin, {\em et~al.}, ``Beta-amyloid deposition and
  glial changes in an {APPPS1} mouse model of {A}lzheimer's disease,'' {\em
  Invest. Ophthalmol. Vis. Sci.} {\bf 57}(12), 2240  (2016).

\bibitem{baumann2019investigating}
B.~Baumann, D.~J. Harper, A.~Lichtenegger, {\em et~al.}, ``Investigating
  retinal changes in a mouse model of {A}lzheimer’s disease using {OCT},''
  {\em Invest. Ophthalmol. Vis. Sci.} {\bf 60}(9), 199  (2019).

\end{thebibliography}
\bibliographystyle{spiejour}   



\end{spacing}
\end{document}